\documentclass[aps, prx, twocolumn,10pt,longbibliography, superscriptaddress]{revtex4-1}

\usepackage{amsmath, amssymb, amsfonts, amsthm, bm, mathrsfs, bbm}
\usepackage{graphicx}
\usepackage{epstopdf}
\usepackage{color}
\usepackage[usenames,dvipsnames]{xcolor}
\usepackage[citecolor=blue, colorlinks=true, urlcolor=blue, linkcolor=blue]{hyperref}

\newcommand{\fig}[1]{Fig.\thinspace{}\ref{#1}}

\newcommand{\fc}[1]{({#1})}
\newcommand{\figc}[2]{Fig.\thinspace{}\ref{#1}\thinspace{}\fc{#2}}

\usepackage{times}

\begin{document}

\title{A self-consistent Hartree-Fock approach to Many-Body Localization}

\author{Simon A. Weidinger}
\email[]{simon.weidinger@tum.de}
\affiliation{Department of Physics and Institute for Advanced Study, Technical University of Munich, 85748 Garching, Germany}

\author{Sarang Gopalakrishnan}
\affiliation{Department of Physics and Astronomy, CUNY College of Staten Island, Staten Island NY 10314 USA and Initiative for the Theoretical Sciences, CUNY Graduate Center, New York, NY 10016 USA}

\author{Michael Knap}
\affiliation{Department of Physics and Institute for Advanced Study, Technical University of Munich, 85748 Garching, Germany}

\begin{abstract}
In this work, we develop a self-consistent Hartree-Fock approach to theoretically study the far-from-equilibrium quantum dynamics of interacting fermions, and apply this approach to explore the onset of many-body localization (MBL) in these systems.
%
We investigate the dynamics of a state with a nonequilibrium density profile; we find that at weak disorder the density profile equilibrates rapidly, whereas for strong disorder it remains frozen on the accessible timescales. We analyze this behavior in terms of the Hartree-Fock self-energy. At weak disorder the self-energy fluctuates strongly and can be interpreted as a self-consistent noise process. By contrast, at strong disorder the self-energy evolves with a few coherent oscillations which cannot delocalize the system. Accordingly, the non-equilibrium site-resolved spectral function shows a broad spectrum at weak disorder and sharp spikes at strong disorder. Our Hartree-Fock theory incorporates spatial fluctuations and rare-region effects. As a consequence, we find subdiffusive relaxation in random systems; but, when the system is subjected to weak quasi-periodic potentials, the subdiffusive response ceases to exist, as rare region effects are absent in this case. This self-consistent Hartree-Fock approach can be regarded as a relatively simple theory that captures much of the MBL phenomenology. 
%

\end{abstract}

\date{\today}

\pacs{
}

\maketitle



\section{Introduction}
\label{sec:Intro}

In recent years the phenomenon of many-body localization (MBL) has attracted major interest, both experimentally~\cite{schreiber2015observation, kondov2015disorder, smith2016many, bordia2016coupling, choi2016exploring, bordia2017periodically, bordia20172D, luschen2017observation, roushanGoogle2017, rosenbaum2017, lukin_probing_2018} and theoretically~\cite{basko2006metal, gornyi2005interacting, imbrie2016many, nandkishore2015many, altman2015universal}. 
MBL systems, unlike generic quantum many-body systems, do not thermalize~\cite{deutsch1991quantum, srednicki1994chaos, rigol2008thermalization, nandkishore2015many}. In these systems, entanglement entropy grows logarithmically, and local quantum correlations survive for long times~\cite{de2006entanglement, ZnidaricPrelovsek08, bardarson2012unbounded, serbyn2014interferometric, bahri2015localization, serbynPapicAbaninQuantumQuenchesPRB201567, torres-herrera_generic_2018, schiulaz_thouless_2018}. 
The transition from the thermal to the MBL phase is an unconventional dynamical phase transition, and its critical properties have attracted much recent attention~\cite{pal, luitz2015many, agarwal2015anomalous, voskhusealtman, pottervasseursid, gopalakrishnan_mean-field_2014-1, dvp, thiery2017, khemani_critical_2017, gornyi_spectral_2017, goremykina18, han_boltzmann_2018}. Since the MBL phase does not thermalize, it is impossible to describe the MBL phase transition and the critical phenomena associated with it in the framework of equilibrium statistical physics. 
The central obstacle is that while the regimes deep in the thermal phase and deep in the localized phase are phenomenologically well understood~\cite{chaikinlubensky, vosk2013many, huse2014phenomenology, serbyn2013local, ros2015integrals}, these phenomenologies (based respectively on equilibrium statistical physics and on local integrals of motion) are incompatible with one another, and both break down as the transition is approached.


In the present paper, we develop a field theoretic description of the many-body localization problem in the two-particle irreducible (2PI) Keldysh framework~\cite{baym1961conservation, keldysh1965diagram, kamenev2011field} by looking at the relaxation dynamics of initial states. Using a self-consistent weak-coupling expansion, we arrive to leading order at a self-consistent Hartree-Fock theory of the many-body dynamics, in which single particles move in the presence of the noise due to the other particles. At the Hartree-Fock level we are able to simulate the dynamics of systems of up to $192$ sites for times up to $10^4/J$ where $J$ is the hopping. The Hartree-Fock theory captures both the slowdown of thermalization and the onset of a delocalized, subdiffusive phase in random systems~\cite{agarwal2015anomalous, voskhusealtman, pottervasseursid, barLev2015, gopalakrishnan2016griffiths, gopalakrishnanMottCond2015, PhysRevB.94.224207, PrelovsekHerbrych17, agarwal2017rare, PhysRevLett.117.040601, Karrasch_2017}.  
By contrast, this subdiffusive phase is absent for systems with quasiperiodic potentials, consistent with Ref.~\cite{znidaric2018}. 
Our approach also gives us access to the non-equilibrium local spectral function, which is expected to look qualitatively different for localized and delocalized systems~\cite{basko2006metal}. Indeed we find, that the nonequilibrium local spectral function shows a broad spectrum at weak disorder, but exhibits sharp spikes at strong disorder (or for strong quasi-periodic potentials). The field-theoretic framework we develop can be extended beyond leading order, although higher orders are numerically intensive.

This paper is organized as follows. In Sec.~\ref{sec:model}, we describe the investigated model as well as the non-equilibrium Keldysh field theory approach we use to calculate the time evolution.
In Sec.~\ref{sec:relax} we show how different initial product states relax due to interaction and that fermion transport is subdiffusive for weak randomness. This is compared to quasi-periodic potentials, for which subdiffusion is absent. We show that the nonequilibrium local spectral function displays a broad spectrum for weak disorder, while it exhibits sharp peaks for strong disorder in Sec.~\ref{sec:spectral}.
In Sec.~\ref{sec:noise} we analyze the amplitude spectrum and the autocorrelations of the Hartree-Fock self-energy and argue that the delocalization of the system can be understood by observing that the Hartree-Fock self-energy acts similar to noise for weak disorder. Finally, we conclude our results in Sec.~\ref{sec:outlook}.


\section{Model and Method}
\label{sec:model}
We study a model of spinless fermions with nearest-neighbor interactions (i.e., a ``spinless Fermi-Hubbard model'') 
\begin{equation}
\hat{H} = -J \sum\limits_{\langle i, j\rangle} \hat{c}_i^\dagger \hat{c}_j + U  \sum\limits_{\langle i, j\rangle} \hat{n}_i \hat{n}_j + \sum\limits_j h_j \hat{n}_j,
\label{eq:FH}
\end{equation}
where $h_i$ are chosen to be either uncorrelated random fields drawn from a box distribution $[-W,W]$ or a quasi-periodically varying potential. 
We quote results for the random case, except when otherwise specified. 
We fix the parameters $J=1$, $U=0.5$ and work with periodic boundary conditions. We investigate the model in one spatial dimension, even though our method can be readily extended to two dimensions.
In one dimension and for the aforementioned set of parameters the Hamiltonian, Eq.~\eqref{eq:FH}, maps via the Jordan-Wigner transformation onto the disordered spin-$1/2$ XXZ-model at $J_z = 0.5 J_\perp$ and a reduced disorder strength $\tilde{W} = 0.5W$. The XXZ-model is a paradigmatic model in the study of many-body localization. Numerical studies based on exact diagonalization indicate a localization-delocalization transition at the critical disorder strength $\tilde{W}_c \simeq 3.6$~\cite{pal, luitz2015many}.

To calculate the time evolution of the system, we use the nonequilibrium Keldysh field theory formalism~\cite{baym1961conservation, keldysh1965diagram, kamenev2011field}. In this approach one propagates the contour ordered Green's functions $G_{ij}(t, t^\prime) = \langle T_\mathcal{C} \hat{c}_i(t) \hat{c}^\dagger_j(t^\prime)\rangle$ on the Schwinger-Keldysh closed time contour (CTC) by solving a nonequilibrium Dyson equation. To make the CTC structure explicit, we introduce lesser and greater Green's functions, $G_{ij}^<(t, t^\prime) = i \langle \hat{c}^\dagger_j(t^\prime) \hat{c}_i(t)\rangle$ and $G_{ij}^>(t, t^\prime) = -i \langle \hat{c}_i(t) \hat{c}^\dagger_j(t^\prime) \rangle$, in which the Dyson equations take the form 
\begin{align}
\left[i \partial_t -\hspace{-1pt} \hat{J} + \hat{\Sigma}^{\mathrm{HF}}(t)\right] \hspace{-3pt}\ast\hspace{0pt} &\hat{G}^\lessgtr(t, t^\prime) = \hspace{-3pt}\int\limits_0^t\hspace{-3pt} dt''~\hat{\Sigma}^{\mathrm{R}}(t, t'') \ast \hat{G}^\lessgtr(t'', t^\prime)\notag\\ +
\int\limits_0^{t'} dt''~\hat{\Sigma}^{\lessgtr}(t, t'') \ast & \hat{G}^\mathrm{A}(t'', t^\prime).
\label{eq:Dyson1}
\end{align}
Here $\ast$ denotes a matrix product over spatial indices $i$, $j$ and $-\hat{J}_{ij} = -J \delta_{\langle i, j\rangle} + h_i \delta_{ij}$ is the sum of the hopping and onsite potential matrices. 
The lefthand side of Eq.\eqref{eq:Dyson1} contains only terms local in time, in particular the Hartree-Fock selfenergy $\Sigma_{ij}^\mathrm{HF}(t)$. The righthand side on the other hand entails integrals over the entire past of the system, which incorporate memory effects in the dynamics. We expect that these time-nonlocal effects are necessary to capture full thermalization at weak disorder; however, as we shall see below, relaxation of a nonequilibrium initial state can be captured even if one neglects memory effects and takes into account only the time-local Hartree-Fock part of the self-energy, $\Sigma_{ij}^\mathrm{HF}(t)$.
The self-consistent Hartree-Fock theory therefore maps Slater determinants to other Slater determinants, and in this sense does not give rise to full thermalization. 

In the following, we focus on the nonequilibrium Dyson equation at the Hartree-Fock level,
\begin{equation}
\left[i \partial_t - \hat{J} + \hat{\Sigma}^{\mathrm{HF}}(t)\right] \ast \hat{G}^\lessgtr(t, t^\prime) = 0.
\label{eq:Dyson2}
\end{equation}
The selfenergy is in general given as the functional derivative of the two particle irreducible (2PI) effective action $\Gamma_2[G]$~\cite{baym1961conservation} with respect to the Green's function and thus is a functional of the full Green's function, in contrast to normal perturbation theory, where one expands the self-energy in the bare Green's function. The Hartree-Fock selfenergy is obtained from an expansion of $\Gamma_2[G]$ to first order in the nearest neighbor repulsion $U$, such that 
\begin{align}
\Sigma_{mn}^\mathrm{HF}(t) &= \Sigma_{mn}^\mathrm{H}(t) + \Sigma_{mn}^\mathrm{F}(t) \notag\\ 
&= 2 U\delta_{mn} \sum\limits_{\langle l, n \rangle} n_l(t) +  2i U \delta_{\langle m, n \rangle} G^<_{mn}(t, t).
\label{eq:HF}
\end{align}
From the lesser and greater Green's functions one can obtain observables like the occupation numbers $n_j(t) = \langle \hat{n}_j(t)\rangle = -i G_{jj}^<(t, t)$ and also the retarded Green's function $G^\mathrm{R}_{ij}(t, t^\prime) = \Theta(t-t^\prime)[G^>_{ij}(t, t^\prime) - G^<_{ij}(t, t^\prime)]$. 

We treat disorder in an exact way by sampling realizations from the disorder distribution, simulating the time evolution of the lesser/greater Green's functions $G^\lessgtr_{ij}(t, t^\prime)$ and in the end averaging the quantity of interest over the samples until convergence is achieved. Typically a few hundred samples are necessary. Therefore, no replica trick~\cite{AndersonReplica}, as usually used in the equilibrium field theoretic treatment of disordered systems, is required.

Due to the absence of memory integrals in Eq.~\eqref{eq:Dyson2} at the Hartree-Fock level, we are able to treat system sizes up to $192$ sites and times up to $10^4$ hopping scales, which thus goes significantly beyond the state of the art of exact diagonalization in system size and matrix product state based approaches in time. Moreover, we show in App.~\ref{sec:hfed}, that our approach captures well the time evolution of small systems calculated by exact diagonalization and discuss field-theoretic results beyond Hartree-Fock in App.~\ref{app:SCBA}. Using a related framework based on quantum master equations that were derived from perturbation theory, Refs.~\cite{barlev1,barlev2} studied the relaxation of disordered fermions. This approach however, does not necessarily conserve energy and the total particle number. The non-equilibrium Dyson equation \eqref{eq:Dyson2} is a first order differential equation in time and one needs to fix initial conditions $G^\lessgtr_{ij}(0, 0)$. In this work we will look at uncorrelated product initial states, which are uniquely defined by the occupations $n_j(0)$. These fix the lesser Green's functions $G^<_{ij}(0, 0) = i \delta_{ij} n_j(0)$ and, using the anti-commutation relations of $\hat{c}_j$,  $\hat{c}^\dagger_j$, also the greater Green's function $G^>_{ij}(0, 0) =-i \delta_{ij}(1 - n_j(0))$.
\section{Results}
\subsection{Non-equilibrium relaxation of initial states}
\label{sec:relax}

\begin{figure*}
	\includegraphics[width=.98\textwidth]{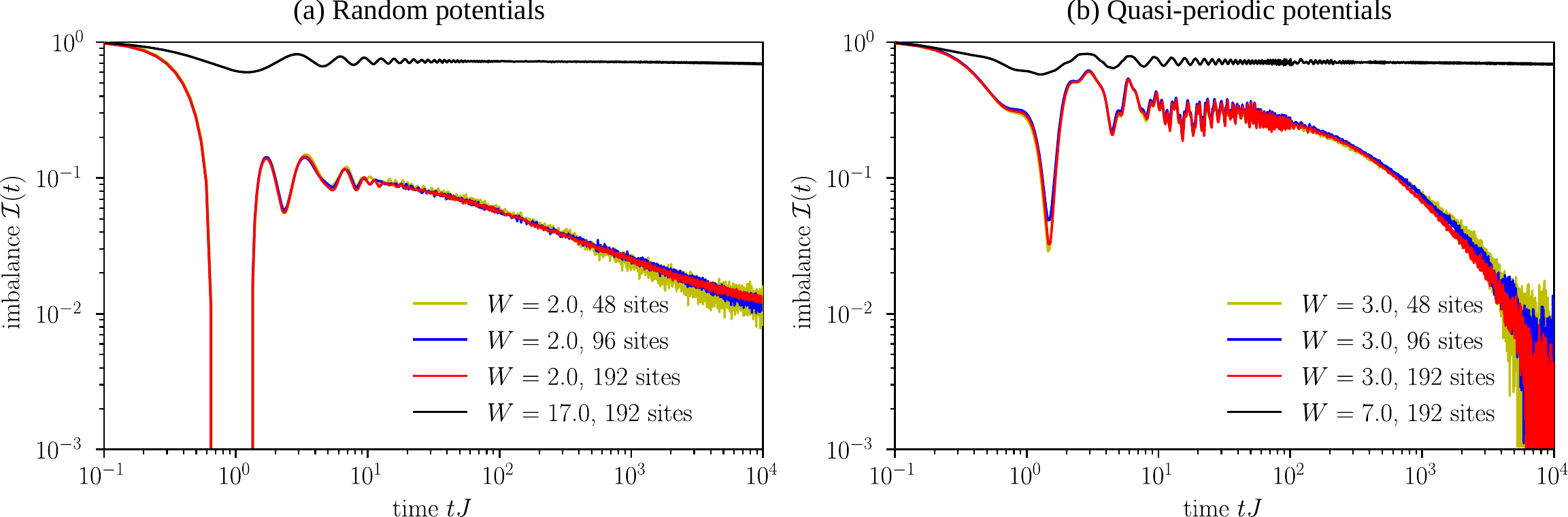}
	\caption{\textbf{Decay of the imbalance for a staggered initial state in one dimension for random and quasi-periodic potentials.} Initially we prepare the system in a staggered product state, where all even sites are occupied and all odd sites are empty. The time evolution of the system is obtained from the Kadanoff-Baym equations including Hartree-Fock effects of a nearest-neighbor repulsion $U=0.5$. \fc{a} In the case of weak disorder, $W=2.0$ (yellow, blue, red), the imbalance $\mathcal{I}=(N_\mathrm{e}-N_\mathrm{o})/(N_\mathrm{e}+N_\mathrm{o})$ decays with a subdiffusive powerlaw, $\mathcal{I}(t)\sim t^{-\alpha}$, with an exponent $0<\alpha<1/2$. Finite size effects are still noticeable for system sizes of $96$ and $192$ sites and become relevant at $t \simeq 5000$. For strong disorder, $W=17.0$, the imbalance $\mathcal{I}$ relaxes to a nonzero value indicating localization of the system (black). Finite size effects are not visible in this case (not shown). \fc{b} For quasi-periodic disorder rare region effects are absent, as the energy mismatches between sites are always either small or large. Hence, in the case of weak disorder, $W=3.0$, there is no subdiffusion and the imbalance $\mathcal{I}(t)$ is decaying faster than a powerlaw (yellow, blue, red).  For strong disorder, $W=7.0$, the system becomes localized (black line).}
	\label{fig:imbalance}
\end{figure*}

The main observable we use to study the relaxation of an initial state, is the density-density correlation function 
\begin{equation}
\mathcal{C}(t) = \frac{2}{N} \sum_j \langle \hat{n}_j(t) \hat{n}_j(0)\rangle -1,
\label{eq:C}
\end{equation}
where $N$ is the number of fermions in the lattice. We consider half filled systems, $N = L/2$, where $L$ is the  size of the lattice. The correlation function has the property that $\mathcal{C}(t=0) = 1$. If the system is localized, i.e., the system remains in a spatially nonuniform state, $\mathcal{C}$ remains finite for all times, $\mathcal{C}(t\rightarrow\infty) \neq 0$. Instead, if the system becomes delocalized and there is a relaxation to a uniform state, the correlation function becomes zero at late times, $\mathcal{C}(t\rightarrow \infty) = 0$~\footnote{This is only strictly valid in the thermodynamic limit, for a finite size system $\mathcal{C}(t)$ will attain a residual value $\sim 1/L$ for late times even in the delocalized phase.}.

As the density-density correlation, Eq.~\eqref{eq:C}, is a four-point function, it is in general not possible to calculate it from two-point Green's functions. However, for product initial states, the four-point function reduces to a two-point function and, the lesser Green's function $G^<_{jj}(t, t)$ is sufficient to obtain $\mathcal{C}(t)$.  
For an initially staggered state, where every other site is occupied, i.e., $n_j(0)=1$ for $j$ even and $n_j(0)=0$ for $j$ odd,
$\mathcal{C}(t)$ is identical to the imbalance $\mathcal{I}(t)=(N_\mathrm{e}-N_\mathrm{o})/(N_\mathrm{e}+N_\mathrm{o})$ between even and odd sites, which is often measured in optical lattice experiments~\cite{schreiber2015observation}.

\subsubsection{Random vs. quasi-periodic potentials}
\label{sec:random}

In order to study random systems, we draw the local potentials from a bounded box-distribution, 
\begin{equation}
h_j \in [-W, W]
\end{equation} 
and refer to that as random disorder. However, recently many experiments have explored MBL using quasi-periodic potentials, 
\begin{equation}
h_j = W\mathrm{cos}(2\pi \Phi j + \theta),
\label{eq:qpdisorder}
\end{equation}
instead of uncorrelated randomness. Here $\Phi$ is the golden ratio and observables are averaged over several values of the phase $\theta$. As the period of the cosine function is incommensurate with the lattice spacing, the potential looks quasi-random. Nevertheless, there are crucial differences to a truly random potential. First of all, there is already a localization transition in the one dimensional non-interacting system, commonly known as the Aubry-Andr\'e model~\cite{aubry1980analyticity}, at $W_c=2$. Secondly, the detuning between neighboring sites is either always small or large throughout the whole system, depending on the value of the strength $W$, and hence rare regions are absent.

We first turn to the case of random disorder potentials. For weak disorder, $W=2.0$, an initially staggered state relaxes and the imbalance decays \emph{in random potentials} as a power law $\mathcal{I}(t)\sim t^{-\alpha}$, with an exponent $\alpha$ between 0 and $1/2$, Fig.~\ref{fig:imbalance}\fc{a}. The powerlaw relaxation occurs due to the presence of rare region/Griffiths effects~\cite{agarwal2015anomalous, gopalakrishnan2016griffiths, agarwal2017rare}. Griffiths effects arise due to the enclosure of non-relaxing localized regions in the otherwise delocalized system. The probability for having a localized inclusion in $d$ dimensions is exponentially small in its size, $\sim q^{l^d}$, and the timescale for such a region to relax is exponentially long, $\sim e^{l/\xi}$, where $\xi$ is the localization length in the inclusion. For uncorrelated disorder, it follows that the density of inclusions that are dynamically frozen on a timescale $t$ is $q^{(\xi \log t)^d}$. In one dimension, these regions hence give rise to a residual contrast that scales as a power-law with a continuously varying exponent as disorder is tuned. In two dimensions, rare regions give rise to a 
faster decay of the contrast. Asymptotically, it is therefore expected that this log-normal decay should be subleading to hydrodynamic long-time tails~\cite{gopalakrishnan2016griffiths}. 

If disorder is strong, $W=17.0$, the system becomes localized and the imbalance $\mathcal{I}(t)$ saturates at a sizable finite value, indicating that the system keeps a memory of the initially imprinted staggered particle distribution. Deep in the localized phase there is no finite size dependence of the imbalance, because the localization length is much smaller than system size and particles do not experience the boundaries of the system. Consequently small size simulations are sufficient to observe the thermodynamic limit. On the other hand, finite size effects are considerable and still relevant even up to the order of hundreds of sites for $W=2$.

\begin{figure}
	\includegraphics[width=.48\textwidth]{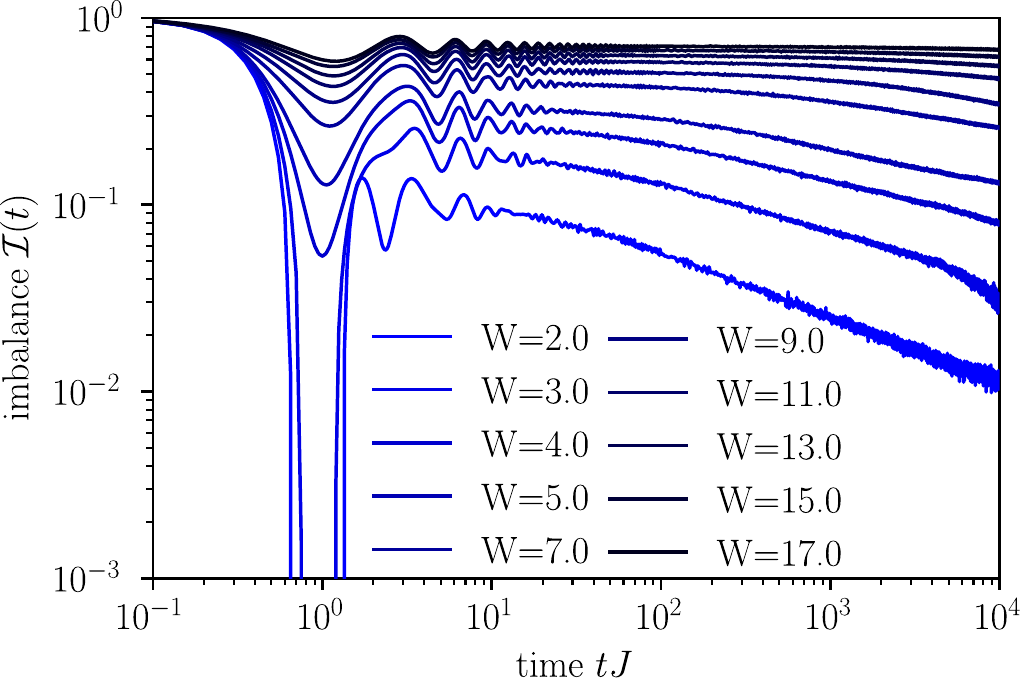}
	\caption{\textbf{Relaxation of the imbalance for varying disorder.} We gradually increase the disorder strength from $W=2$ to $W=17$ for systems of 192 sites, and monitor the relaxation dynamics of the imbalance $I(t)$ in time. Within the self-consistent Hartree-Fock approach we find that the system keeps its memory of the initial state to late times for disorder strength $W \gtrsim 15$, which corresponds to a disorder strength of $\tilde{W}\gtrsim7.5$ in the XXZ-model.}
	\label{fig:WComparison}
\end{figure}
\begin{figure*}
	\includegraphics[width=.98\textwidth]{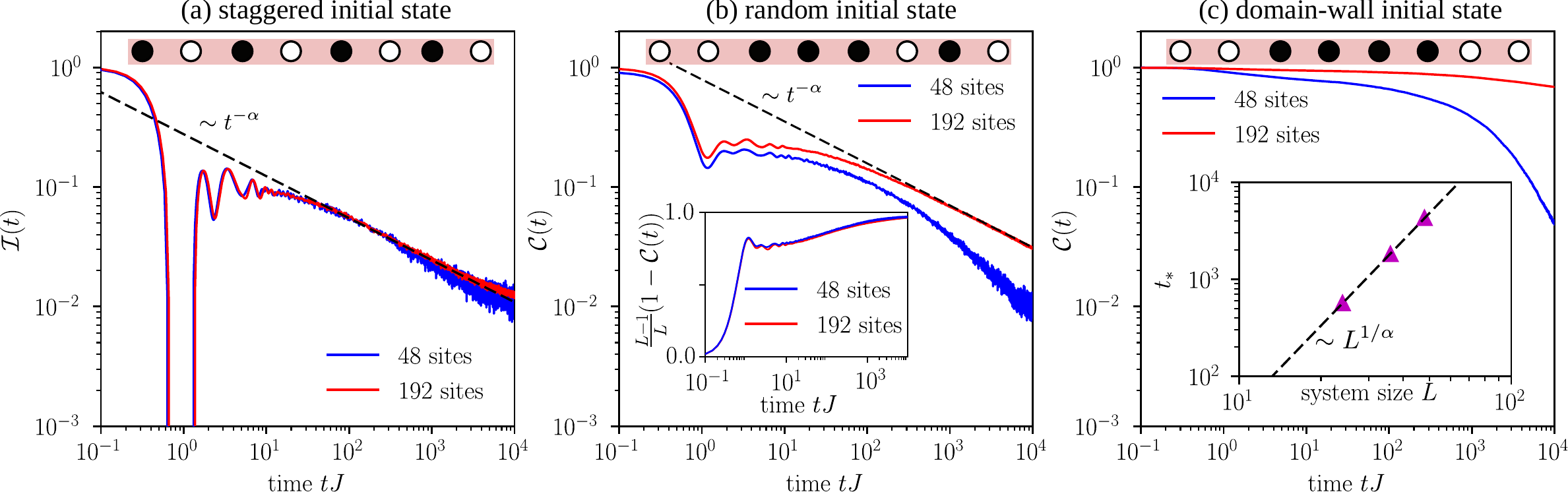}
	\caption{\textbf{Subdiffusive decay of different initial states for random disorder potentials.} We prepare the system in three different initial states, in a staggered initial state \fc{a} where every other site is occupied, in a random initial state \fc{b} where half of the sites are occupied at random positions, and in a domain-wall state \fc{c} where a chain of length $L/2$ in the middle of the system is occupied with fermions. The states are depicted schematically at the top of each panel. 
	In the case of a staggered initial state the density-density correlation function $\mathcal{C}(t)$ reduces to the imbalance $\mathcal{I}(t)$. \fc{a} The staggered initial state shows the fastest relaxation and the imbalance decays according to a subdiffusive powerlaw $\mathcal{I}(t)\sim t^{-\alpha}$. (For the disorder strength shown, $\alpha=0.351$). \fc{b} Relaxation is slower for random initial states as they typically contain small blocks of occupied sites, which relax slowly due to the Pauli principle. For late times however the density-density correlation approaches the same subdiffusive powerlaw $\mathcal{C}(t) \sim t^{-\alpha}$ as the staggered initial state. To see this, large system sizes of several hundreds sites are required. The inset shows the proper short time finite size scaling of the correlation function. \fc{c} The domain-wall initial state is slowest to relax due to the Pauli principle blocking hopping inside the block. The timescale $t_\ast$ for melting the block, reminiscent of the Thouless time in diffusive systems, shows the scaling $t_\ast \sim L^{1/\alpha}$ with system size (inset), consistent with the presence of subdiffusive transport in the system. All graphs are shown for nearest neighbor repulsion $U=0.5$ and disorder strength $W=2.0$.}
	\label{fig:relax}
\end{figure*}

The relaxation of the imbalance for various values of the disorder $W$ and systems of 192 sites is shown in \fig{fig:WComparison}. Within the self-consistent Hartree-Fock theory, the imbalance decays for a large parameter regime to zero with a subdiffusive power law. Only for $W \gtrsim 15$ (corresponding to $\tilde{W}\gtrsim 7.5$ in the XXZ-model), the imbalance ceases to relax on the simulated times.  We argue in Sec.~\ref{sec:noise} for the existence of a true localization transition within our approximations, which would thus be at significantly larger disorder strength than the one obtained from small scale numerics, which has been estimated to be at $\tilde{W} \sim 3.6$ for our system~\cite{pal, ZnidaricPrelovsek08, bardarson2012unbounded, luitz2015many}. Note that previous large-system studies using the numerical linked-cluster approach~\cite{devakul2015} and a recent study using matrix product states~\cite{2018arXiv180705051D} had also located the transition at much stronger disorder.


With quasi-periodic potentials we again find that including interactions at the Hartree-Fock level leads to a relaxation of the initially imprinted density pattern for weak disorder, $W=3.0$, Fig.\ref{fig:imbalance}\fc{b}. Yet, due to the strong spatial correlations of the quasi-periodic disorder, rare region effects are absent and our numerics does not show subdiffusive power-law decay of the imbalance. This is in contrast to a tDMRG and renormalization group study, which is however limited to shorter time scales~\cite{Karrasch_2017}. Due to the fast decay one can still see the residual finite size value of $\mathcal{I}(t)$ even for system sizes of $192$ sites at late times. 

Note, that in the absence of interactions, the localization length $\xi\sim1.2$ of the model with quasi-periodic disorder at $W=3.0$ is significantly shorter than that of the model with random disorder at $W=2.0$ ($\xi\sim4.0$). Yet the imbalance decays much faster for quasi-periodic disorder, therefor we can expect, that the absence of a subdiffusive powerlaw decay is a genuine effect of the type of the disorder potential, rather than a mere disorder strength effect.

Increasing the disorder strength, the system localizes at weaker potential strength than in the case of a true random potential; it is already fully localized for $W=7.0$. This is consistent with the intuitive picture of energy mismatches between sites becoming large everywhere without any statistical fluctuations.


\subsubsection{Initial-state dependence}
\label{sec:initstates}

Besides the staggered state we also consider random initial states, where the initially occupied sites are chosen randomly as well as a domain-wall initial state where a block of $L/2$ sites in the middle of the system is initially occupied. 
In Fig.~\ref{fig:relax} we compare for \emph{random disorder} the relaxation of the three type of initial states (staggered, random, and domain wall). 
From a coarse-grained, hydrodynamic point of view, these three types of initial states differ in that the staggered state is dominated by high-momentum fluctuations, the domain-wall initial state has exclusively low-momentum fluctuations, and the random state has fluctuations at all scales. As expected on general hydrodynamic grounds, therefore, the domain-wall state is much slower to relax than the staggered state: the timescale on which it relaxes can be interpreted as the Thouless time for the system. 

The decay time scales of random initial states, Fig.~\ref{fig:relax}\fc{b}, is in between the staggered initial state, \fc{a} and the block initial state, \fc{c}. For random initial states, the density-density correlation $\mathcal{C}(t)$ is not only averaged over disorder realizations but also over different initial particle distributions, such that this case can be interpreted as the result for an infinite temperature ensemble. For large system sizes of $192$ sites, the decay of $\mathcal{C}(t)$ approaches the same power-law decay $\sim t^{-\alpha}$ as the imbalance in the case of a staggered initial state at late times. This behavior is not observable for smaller systems, which in particular implies that numerics for small system sizes is not sufficient to study the power law relaxation in that observable.

To further corroborate the observation of subdiffusive particle transport in the system at weak disorder, we analyze the finite size scaling of the crossover timescale $t_\ast$, at which the block initial state starts to decay, see Fig.~\ref{fig:relax}\fc{c} (inset). For concreteness we define $t_\star$ as the time where $\mathcal{C}(t)$ has dropped to $1/(2e)$, though we have checked, that the scaling is insensitive to this specific choice. In a diffusive system the notion of $t_\star$ would be equivalent to the Thouless time $t_\mathrm{Th}$, which scales quadratically with system size, $t_\mathrm{Th}\sim L^2$. For subdiffusive transport we find a steeper power law $t_\ast\sim L^{1/\alpha}$, where $\alpha$ is the exponent of the imbalance decay. The two exponents coincide, at least within the Hartree-Fock theory, because the timescale governing density relaxation across the system is the relaxation timescale of the slowest bottleneck expected in a system of size $L$. This slowest bottleneck should correspond to a Griffiths region with density $1/L$. Given that the density of a Griffiths region with timescale $t$ scales as $1/t^\alpha$, we thus have that $t_\ast \sim L^{1/\alpha}$ (dashed line in the inset of \figc{fig:relax}{c}). Our numerics yields a very good agreement of  these two exponents, see Fig.~\ref{fig:relax}\fc{a} and \fc{c}


Looking at panels \fc{b} and \fc{c} of Fig.~\ref{fig:relax}, it can be seen, that finite size effects become noticeable already at $tJ\sim 1$ for a random or block initial state, while they only become relevant at late times for a staggered initial state, Fig.~\ref{fig:relax}\fc{a}. This feature follows from the definition of $\mathcal{C}(t)$: for a staggered initial state, local relaxation takes place everywhere in the system at once, so the system-wide imbalance drops at a rate of order unity; but for a domain-wall initial state, all the relaxation takes place at the boundary, so the system-wide imbalance drops at a rate $O(1/L)$. The random initial state is, again, in between the two cases. Calculating the average number of domain walls for a chain of length $L$ and periodic boundary conditions, as $\frac{L}{2}\big(1+\frac{L-4(N-L/2)^2}{L(L-1)}\big)$, we can remove the short time finite size effect, by properly rescaling the measured correlation function $\mathcal{C}(t)$, see inset of Fig.~\ref{fig:relax}(b).


\begin{figure*}
	\includegraphics[width=.98\textwidth]{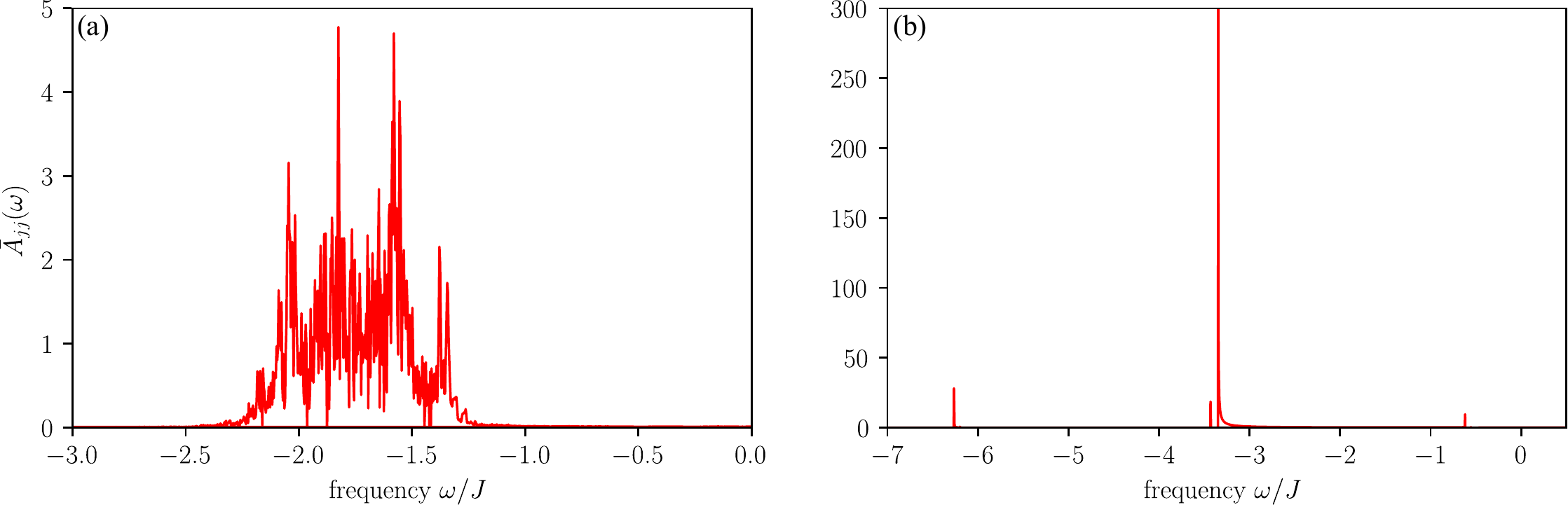}
	\caption{\textbf{Nonequilibrium local spectral function for a single disorder realization.} Averaging the nonequilibrium spectral function $A_{jj}(T, \omega)$ over the center-of-mass time $T$ extracts the non-negative, non-oscillatory part $\bar{A}_{jj}(\omega)$, which can be interpreted analogously to the usual equilibrium spectral function. \fc{a} For weak disorder, $W=2.0$, the local spectral function $\bar{A}_{jj}(\omega)$ has a broad support and hence there is a finite energy window in which fermions can tunnel into/out of the site. This leads to the delocalization of the system. \fc{b} In the case of strong disorder, $W=17.0$, $\bar{A}_{jj}(\omega)$ has sharp delta spikes only for a finite set of frequencies, thus only fermions with energies from a set of measure zero can hop into or out of the site. Accordingly the system becomes localized. It is crucial to look at single sites and a single disorder realization, as averaging over either lattice sites or disorder would yield a sum of delta peaks which smears out the spectral function.}
	\label{fig:spectrum}
\end{figure*}

\subsection{Spectral information}
\label{sec:spectral}
A useful quantity in the study of many-body localization is the local spectral function, defined as the imaginary part of the retarded Green's function, $A_{jj}(T, \omega) = -1/\pi~\mathrm{Im}G^\mathrm{R}_{jj}(T, \omega)$. Here, $T=(t + t')/2$ is the so called center-of-mass time and the Fourier transform to frequency space $\omega$ is calculated with respect to the relative time $t_\mathrm{rel}=t-t'$. In a full nonequilibrium setting, Green's functions do not only depend on the time difference $t_\mathrm{rel}$ but also on the absolute center-of-mass time $T$. Nonetheless, we will show in the following, that one can still extract similar information as in equilibrium. 

Using the Lehmann representation assuming a nonequilibrium initial state, $A_{jj}(T, \omega)$ can be decomposed as
\begin{align}
&A_{jj}(T, \omega) = \bar{A}_{jj}(\omega) + R_{jj}(T, \omega)\notag\\
&\bar{A}_{jj}(\omega) = \sum\limits_{n, m} |\psi_n|^2 \{|C^{(j)}_{nm}|^2 \delta(\omega- \epsilon_{nm}) + n \leftrightarrow m \},
\label{eq:Adecomp}
\end{align}
where $\psi_n = \langle n|\psi \rangle$, $C^{(j)}_{nm} = \langle n|\hat{c}_j|m \rangle$, $|n\rangle$ are the exact many-body eigenstates of the system, and $\epsilon_{nm}$ are the levelspacings between eigenenergies. The second term in the decomposition, 
\begin{align}
R_{jj}(T, \omega) = &\sum\limits_{m,  n\neq l} \biggl\{ \mathrm{Re}[a_{mnl}^{(j)}(T)] \delta\left(\omega- E_m + \frac{E_n + E_l}{2}\right)\notag\\ 
&+\mathrm{Re}[a_{mln}^{(j)}(T)] \delta\left(\omega+ E_m - \frac{E_n + E_l}{2}\right)\notag\\
&+\frac{1}{\pi}\mathrm{Im}[a_{mnl}^{(j)}(T)] \frac{1}{\omega - E_m + \frac{E_n + E_l}{2}}\notag\\
&-\frac{1}{\pi}\mathrm{Im}[a_{mln}^{(j)}(T)] \frac{1}{\omega + E_m - \frac{E_n + E_l}{2}}\biggr\},
\label{eq:R}
\end{align}
contains all the dependence on center-of-mass time via the time-dependent coefficients $a_{mnl}^{(j)}(T)=e^{-i \epsilon_{nl}T}\psi_n^\ast \psi_l C_{nm}^{(j)} C_{lm}^{(j)\ast} $. Due to these oscillatory contributions, $R_{jj}(T, \omega)$ can become negative, which invalidates the positivity sum rule of the equilibrium spectral function. In contrast to the equilibrium spectral function, out of equilibrium $A_{jj}(T, \omega)$ also shows $1/\omega$ divergences due to the contribution of $R_{jj}(T, \omega)$, which would appear only in the real part of an equilibrium Green's function. 
By contrast, $\bar{A}_{jj}(\omega)$, which we will refer to as local spectral function in the following, has a form similar to an equilibrium spectral function. It is independent of the center-of-mass time $T$, nonnegative, and a weighted sum of $\delta$-functions located at spectral lines $E_n - E_m$ of the system. One can obtain $\bar{A}_{jj}(\omega)$ from $A_{jj}(T, \omega)$ by averaging over center-of-mass time, $\bar{A}_{jj}(\omega) = \lim_{T\rightarrow\infty} T^{-1} \int_0^T dS~A_{jj}(S, \omega)$ as the average will cancel the oscillatory terms in $ A_{jj}(T, \omega)$.

Physically the local spectral function $\bar{A}_{jj}(\omega)$, is interpreted as the amplitude for a fermion with energy $\omega$ tunneling into or out of lattice site $j$. If disorder is weak and the system is delocalized, the local spectral function is finite for a continuous set of frequencies such that fermions with many different energies are able to tunnel into or out of a given site, see Fig.~\ref{fig:spectrum}\fc{a}. In contrast, in the localized phase hopping into or out of a given site is only possible for fermions with a discrete set of energies, hence the local spectral function has only discrete sharp spectral lines, see Fig.~\ref{fig:spectrum}\fc{b}.

As one is limited to a finite time evolution in numerics we average over center-of-mass times in the range $1000 \leq TJ \leq 7313$. Despite this already large averaging window, there are still some artifacts, like zero-crossings and $1/\omega$-singularities, of the $T$-dependent part $ A_{jj}(T, \omega)$ visible in Fig.~\ref{fig:spectrum}\fc{b}. This is due to the presence of oscillations with very long period, or in other words very close energy levels, which naturally appear in localized systems due to the absence of level repulsion.

\begin{figure*}
	\includegraphics[width=.98\textwidth]{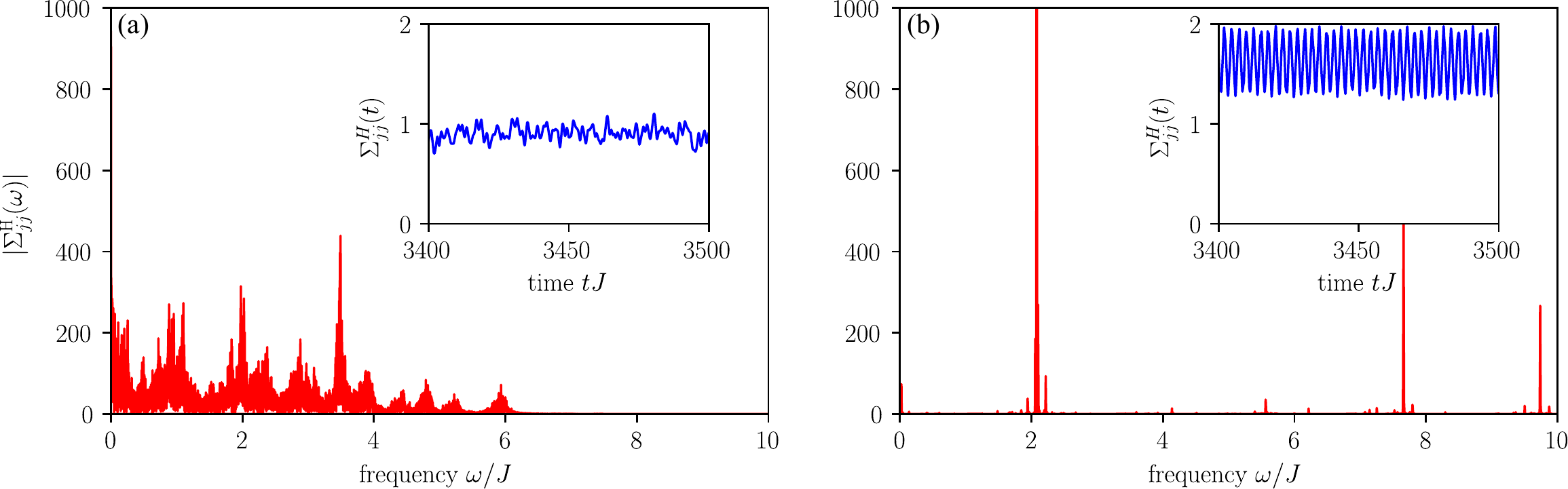}
	\caption{\textbf{Self-consistent noise interpretation of the Hartree selfenergy.} Delocalization in our system on the Hartree-Fock level can be understood by interpreting the Hartree-Fock self-energy $\Sigma_{ij}^\mathrm{HF}(t)$  as noise. Since $\Sigma_{ij}^\mathrm{HF}(t)$ is a functional of the Green's function, the noise is self-consistently generated by the system itself and hence it does not necessarily lead to relaxation. 
	\fc{a} The amplitude spectrum $|\Sigma_{jj}^\mathrm{H}(\omega)|$ of the Hartree selfenergy shows mixing of a continuous frequency range without any to prominent features for weak disorder, $W=2.0$. In the time domain, $\Sigma_{jj}^\mathrm{H}(t)$ therefor looks like noise (inset) and consequently the system delocalizes. On the other hand, for strong disorder $W=17.0$ \fc{b}, the amplitude spectrum only shows mixing of a discrete number of frequencies. Thus, $\Sigma_{jj}^\mathrm{H}(t)$ oscillates coherently in the time domain (inset) and the system remains localized. All data is shown for systems of size L=96.}
	\label{fig:SEnoise}
\end{figure*}

\section{A self-consistent noise interpretation}
\label{sec:noise}

It is often assumed, that treating interactions at Hartree-Fock level is not sufficient to witness the breakdown of localization in a disordered system~\cite{basko2006metal}. While this is true in thermal equilibrium, as we show in Sec.~\ref{sec:relax} this assumption does not hold in the case of quench dynamics that is considered here. 
The essential mechanism by which Hartree-Fock terms cause delocalization is as follows: the time-dependent Hartree-Fock potentials act as effective temporal noise, and a noninteracting system subject to noise will thermalize (though potentially with transient subdiffusive dynamics~\cite{gopalakrishnan2017noise}). 
Intuitively, this can be understood as the noise process, with its continuous frequency spectrum, providing the missing energy for a fermion to hop between two energy-detuned sites. 
We also show in App.~\ref{sec:single}, that relaxation at weak randomness is not due to dephasing effects between different disorder samples or due to averaging over different lattice sites, by computing the relaxation of the local density at single sites and for single disorder configurations.



In our Hartree-Fock theory the selfenergy $\Sigma_{ij}^\mathrm{HF}(t)$ is a deterministic function for a given disorder sample and a given initial state. However, its frequency spectrum can still potentially be broad, allowing for transitions between energy-detuned single-particle orbitals. To analyze this we compute the amplitude spectrum of the Hartree-Fock selfenergy $|\Sigma_{ij}^\mathrm{HF}(\omega)|$ for a single realization of the disorder. As the results are very similar for both the Hartree and the Fock contribution (see App.~\ref{sec:hfself}), we will focus on the amplitude spectrum of the Hartree selfenergy $|\Sigma_{jj}^\mathrm{H}(\omega)|$ only. If the amplitude spectrum is broad and mostly featureless as for disorder strength $W=2.0$ in Fig.~\ref{fig:SEnoise}\fc{a},  the self-energy will look like noise in time domain, leading to delocalization (inset). When the amplitude spectrum consists only of a discrete set of sharp peaks, as is the case for strong disorder $W=17.0$, Fig.~\ref{fig:SEnoise} (b), the Hartree selfenergy is just a coherent oscillation in time (inset), leaving localization intact.

These numerical findings are consistent with what one might expect perturbatively, at weak interactions. A single lattice site overlaps with $\sim \xi$ single-particle orbitals, each at a different energy, and therefore the on-site potential fluctuates at $\sim \xi$ separate oscillation frequencies. At the same time, the energy detuning between a particular orbital and the others it overlaps with goes as $\delta_\xi \sim 1/\xi^2$ (or, more generally, polynomially in $1/\xi$). Thus, when $\xi \gg 1$, a particle in a given orbital is driven at enough different frequencies that it is likely to find a ``noise''-induced resonant transition to another orbital. These transitions lead to yet more frequencies in the self-energy spectrum, inducing yet more transitions, and so forth, and eventually all particles delocalize. In the opposite limit, $\xi \ll 1$, the same logic indicates that localization is stable. In that limit, the amplitude of the Hartree self-energy at a typical site falls off as $\exp(-1/\xi)$ (from orbitals centered at nearest-neighbor sites; further orbitals are exponentially suppressed as $\exp(-L/\xi)$~\cite{nandkishore2014spectral}). For the same reason, a typical orbital has exponentially weak matrix elements to couple to any other orbitals. Thus, asymptotically, in this limit each orbital is subject to a weak, essentially time-periodic potential, which does not induce resonances, leading to a stable localized regime within this approximation.

\begin{figure}
	\includegraphics[width=.46\textwidth]{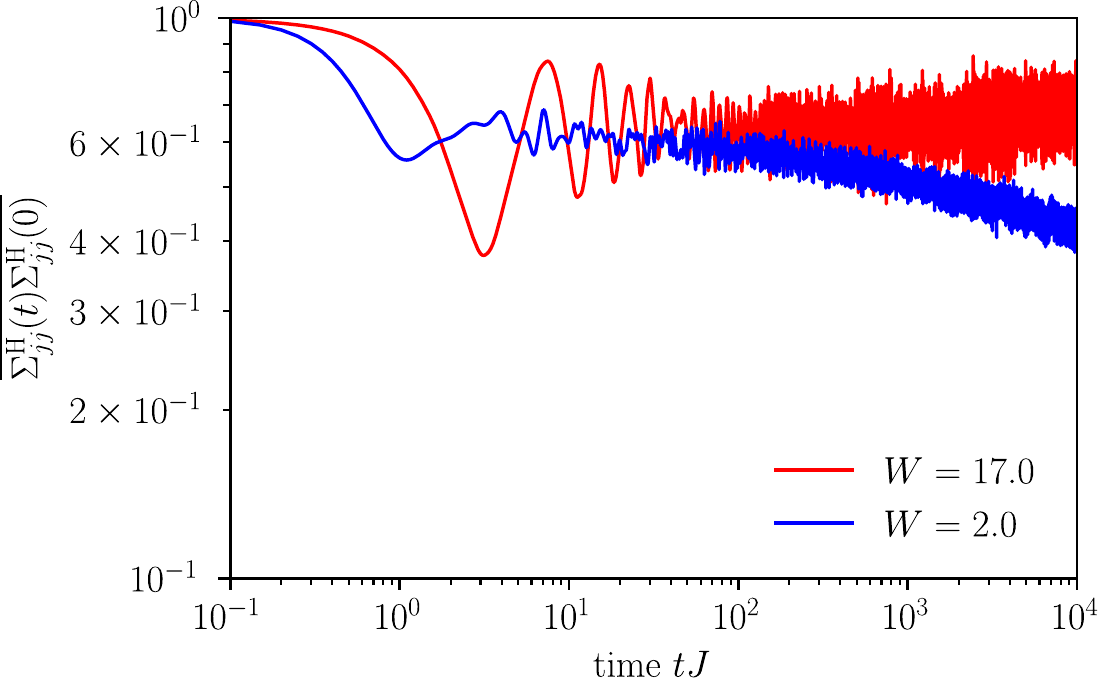}
	\caption{\textbf{Autocorrelation of the Hartree selfenergy.} We compute the autocorrelation $\overline{\Sigma_{jj}^\mathrm{H}(t) \Sigma_{jj}^\mathrm{H}(0)}$ of the Hartree selfenergy averaged over disorder realizations for systems of size L=96. The autocorrelation of $\Sigma_{jj}^\mathrm{H}(t)$ shows a decay in time, consistent with a slow powerlaw, when the disorder is weak, $W=2.0$ (blue). One can therefore think of $\Sigma_{jj}^\mathrm{H}(t)$ as powerlaw correlated noise in a zeroth order approximation. In the localized phase, $W=17.0$, the Hartree selfenergy shows non-decaying autocorrelations for all times (red).}
	\label{fig:SEcorr}
\end{figure}

We now calculate the autocorrelation function $\overline{\Sigma_{jj}^\mathrm{H}(t) \Sigma_{jj}^\mathrm{H}(0)}$ via averaging over disorder realizations, Fig.~\ref{fig:SEcorr}. We find that the autocorrelations show a decay which is consistent with a slow powerlaw $\overline{\Sigma_{jj}^\mathrm{H}(t) \Sigma_{jj}^\mathrm{H}(0)}\sim t^{-\beta}$ in the delocalized phase, $W=2.0$. These noise correlations are much longer lived than in Ref.~\cite{gopalakrishnan2017noise}, where exponential correlations have been studied. However, when one repeats the arguments therein with power-law correlated noise, it turns out that the distribution of tunneling times is still fat tailed and hence subdiffusive transport is recovered in the system, which is consistent with our numerical observations. In contrast, for strong disorder, $W=17.0$, the self-energy autocorrelations remain constant in time.

Despite the similarities between our Hartree-Fock treatment and Ref.~\cite{gopalakrishnan2017noise}, there are still some differences. Most importantly, the noise is generated self-consistently, so its strength is in general not constant in time. The distribution of $\Sigma_{ij}^\mathrm{HF}(t)$, obtained from disorder sampling is furthermore non-Gaussian. We expect that these differences lead to quantitative changes in the dynamics, which need to be addressed in more detailed, future investigations.

\section{Conclusion and Outlook \label{sec:outlook}}
In this work, we developed a self-consistent Hartree-Fock approach in the framework of the nonequilibrium Keldysh field theory to study interacting and disordered fermions, initialized in a far-from-equilibrium state. Our results show that this approach can capture a lot of the phenomenology of many-body localization. Using this technique, we study the time-evolution of systems of up to $192$ lattice sites to times $10^4/J$. With that we can treat systems that are much larger than the ones accessible in exact diagonalization and study dynamics to times much longer than the ones accessible with matrix product states. Moreover, our results also indicate that near the many-body localization transition finite size effects are strong, and for certain observables systems of several hundreds of sites are required to obtain the asymptotic behavior. 

The self-consistent Hartree-Fock approach is sensitive to rare-regions and therefore captures subdiffusive transport for weak random disorder. We showed that in correlated quasi-periodic potentials such subdiffusive transport does not exist, as can be understood from the absence of rare regions. The delocalization of our system on the Hartree-Fock level for weak disorder results from the dynamical nature of the self-energy, which we interpret as noise. For strong disorder, only a couple of frequencies contribute to the self-energy, and hence localization persists. 

From a certain perspective it is surprising that Hartree-Fock theory is able to capture so much of the MBL phenomenology. This appears to have to do with our far-from-equilibrium initial state, which (together with the randomness) builds in fluctuations at many frequencies into the initial conditions for the Hartree-Fock dynamics. If we had instead started with an eigenstate of the noninteracting problem~\cite{basko2006metal, barlev1} the Hartree-Fock theory would not give rise to thermalization, at least for weak interactions. The fact that the performance of mean-field approaches is sensitive to the fluctuations encoded in the initial state---as we see here---was recently pointed out in Ref.~\cite{WURTZ2018341}.

For future work, it will be interesting to study many-body localization in higher dimension and the effects of long-range interactions on the many-body localization transition with this approach and with that explore many-body localization in trapped ions~\cite{smith2016many}, polar molecules~\cite{moses_new_2017}, or condensed matter systems with dipolar interactions~\cite{rosenbaum2017}. Moreover, many-body localized systems that are subjected to periodic driving fields~\cite{bordia2017periodically} can be explored as well with this technique for large system sizes. From a more fundamental point of view, it would intriguing to investigate how one can use measures of the ``spikiness'' of the local spectral function to quantify the many-body localization transition, how they can be measured in ARPES type experiments for ultracold atoms~\cite{bohrdt_angle-resolved_2018}, and whether these measures are consistent with the long-time evolution of the system.

\begin{acknowledgments}
We acknowledge support from the Technical University of Munich - Institute for Advanced Study, funded by the German Excellence Initiative and the European Union FP7 under grant agreement 291763, from the DFG grant No. KN 1254/1-1, DFG TRR80 (Project F8), and the NSF under Grant No. DMR-1653271. S.G. acknowledges the hospitality of the Aspen Center for Physics, which is supported by National Science Foundation grant PHY-1607611.
\end{acknowledgments}

\appendix

\section{Comparison of Hartree-Fock with exact diagonalization \label{sec:hfed}}

\begin{figure}
	\includegraphics[width=.46\textwidth]{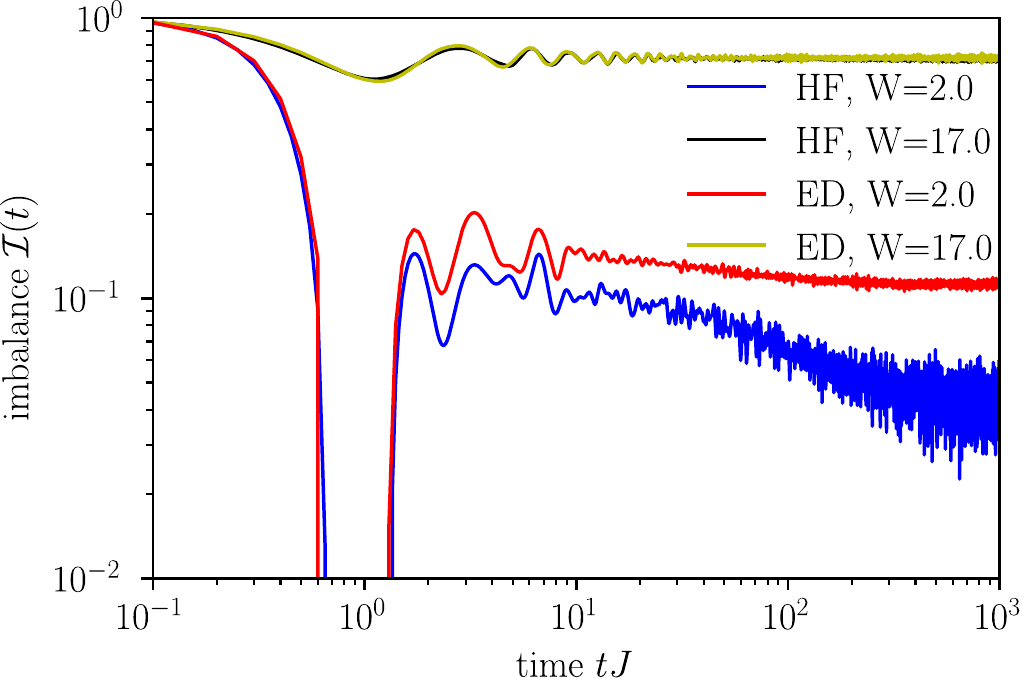}
	\caption{\textbf{Comparison between Hartree-Fock and exact diagonalization}. We compare the Hartree-Fock (HF) theory with exact diagonalization (ED) for small systems of 12 sites at weak disorder, $W=2.0$, where HF delocalizes faster than ED, and at strong disorder, $W=17.0$, where both are practically lying on top of each other. The faster decay of the HF time trace compared to ED for weak disorder can be first attributed to the fact that interactions are treated only perturbatively and second because field theories are not very sensitive to finite size effects. }
	\label{fig:ED} 
\end{figure}

For small systems of $12$ sites we compare the imbalance time-traces obtained from our Hartree-Fock approach to exact-diagonalization calculations in order to get a quantitative benchmark for our method. 
In the weak-disorder regime, Hartree-Fock tends to be more delocalizing, see \fig{fig:ED}. In the strong disorder limit, however, our method is in good quantitative agreement with exact diagonalization. The deviations at weak disorder stem on the one hand from neglecting higher order contributions to the self-energy and on the other hand from the fact that field theories have the tendency to mimic larger systems and hence provide results that are more delocalized compared to exact diagonalization. 

\section{Self-consistent Born approximation \label{app:SCBA}}
\begin{figure}
	\includegraphics[width=.46\textwidth]{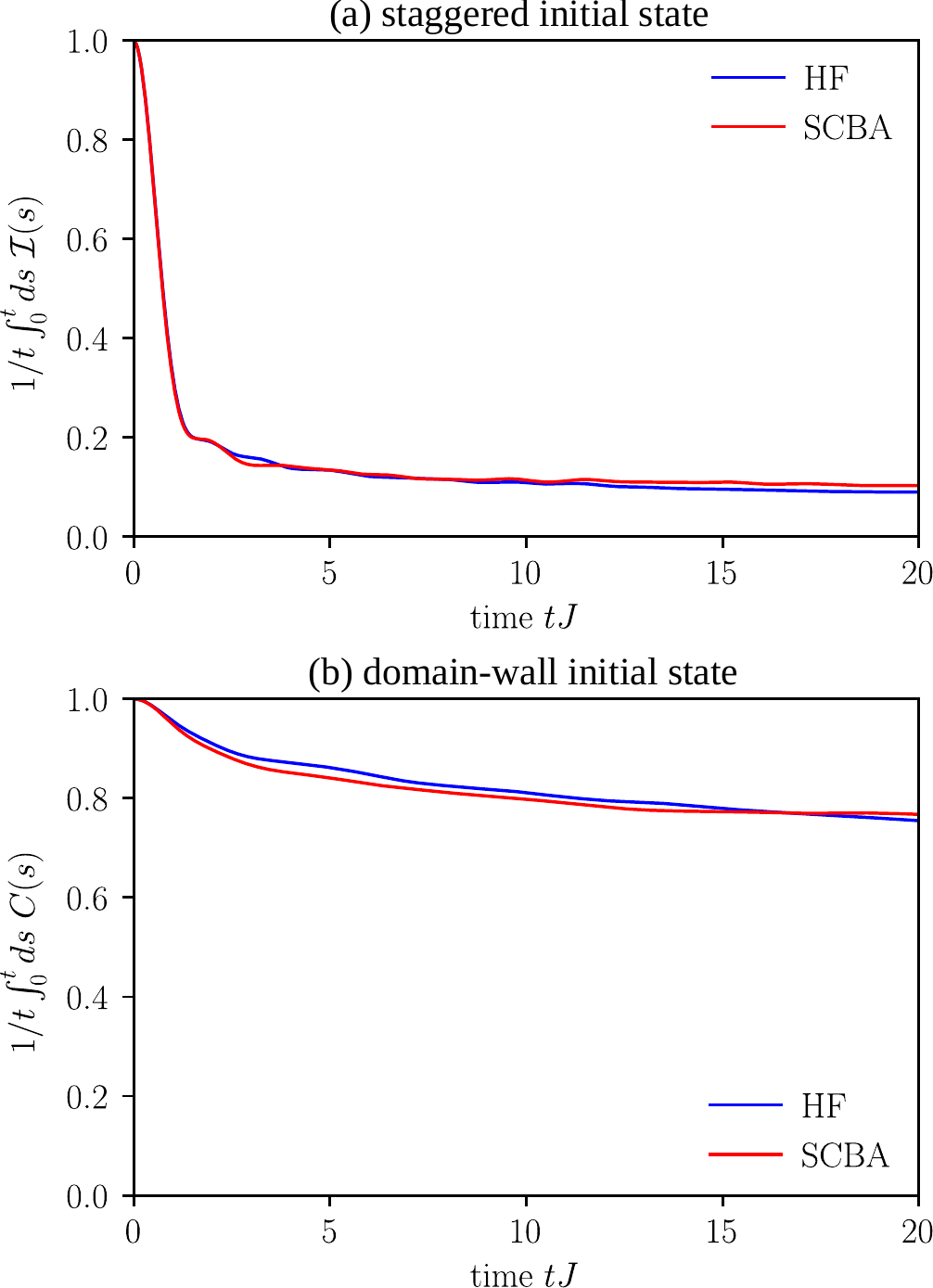}
	\caption{\textbf{Comparison of Hartree-Fock and SCBA at short times.} We compare the leading order Hartree-Fock dynamics with the nex-to-leading order self-consistent Born approximation, both for the (a) staggered and the (b) domain-wall initial state. The data shown is for a single disorder realization and a system size of $L=48$}
	\label{fig:HF+SCBA} 
\end{figure}
 
The next-to-leading order contribution to the self-energy in our weak coupling expansion is often referred to as self-consistent Born approximation (SCBA), which is of second order in the interaction $U$. We calculate the SCBA contribution to the selfenergy and obtain
\begin{equation}
\Sigma_{ij}^\mathrm{SCB}(t, t') = 8 G_{ij}(t, t') \sum\limits_{lk} U_{ik}U_{lj} G_{kl}(t, t') G_{kl}(t', t) .
\label{eq:SCBA_selfenergy}
\end{equation}
We evaluate the Dyson equation taking self-energy contribution up to the SCBA and compute the time averaged correlation function for a staggered initial state and a domain wall initial state, see Fig.~\ref{fig:HF+SCBA}. Within the SCBA it is numerically expensive to reach late times, because the memory integrals on the right hand side of the Dyson equation \eqref{eq:Dyson1} have to be computed. Therefore, our data is limited to times $tJ \sim 20$.

Hartree-Fock is overall consistent with SCBA, but the latter potentially delocalizes the system slightly less, at least on the accessible time scales. On the one hand, SCBA adds new decay channels to the dynamics which on the first sight should enhance delocalization, but on the other hand it might also decrease the self-consistent noise because the memory integral in the Dyson equation damps oscillations.



One possible scenario could be, that the second effect (weaker noise due to damping of oscillations) dominates at short times, before the first effect (larger number of decay channels) takes over at later times, as the SCBA contributions in Eq.~\eqref{eq:Dyson1} may build up slowly over time. Similar behavior has been found in the NLO dynamics of the $O(N)$~\cite{AartsBerges01, BERGES2001369, BERGES2002847, Weidinger17}.

\section{Single samples and single sites \label{sec:single}}

It is well established, that the nonequilibrium Keldysh 2PI approach is able to describe thermalization in a closed system. This is, however, only true, when the memory integrals on the right hand side of Eq.~\eqref{eq:Dyson1}, i.e., the time non-local parts of the selfenergy, are included. The Hartree-Fock selfenergy on its own does not lead to thermalization. Nevertheless we have shown in our present work, that a pure Hartree-Fock time-evolution is already able to describe \emph{relaxation} due to interaction effects in a disordered system. 
\begin{figure}
	\includegraphics[width=.46\textwidth]{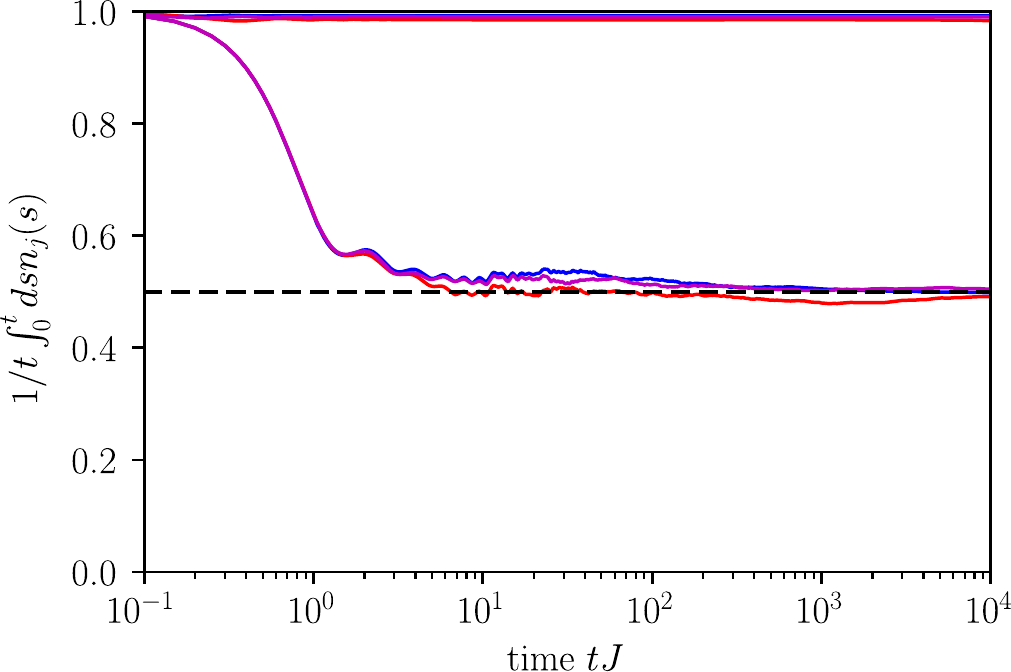}
	\caption{\textbf{Relaxation of an initially occupied site for a single disorder realization.} To verify that the decay of the density-density correlation is not due to dephasing or averaging effects, we compute the time averaged occupation on a single site and for a single disorder realization. For weak disorder, $W=2.0$, (lower curves) this quantity decays to the average density of a half-filled lattice. By contrast, for strong disorder, $W= 17.0$, Without time averaging, $n_j(t)$ would be persistently oscillating about $1/2$ at late times. Therefor the decay of the imbalance $\mathcal{I}(t)$ or in general the correlation function $\mathcal{C}(t)$ is due to particle transport. For strong disorder, $W=17.0$, the particle number does not decay on an initially occupied site, indicating localization of particles on the given site.}
	\label{fig:singlesite}
\end{figure}
\begin{figure}
	\includegraphics[width=.46\textwidth]{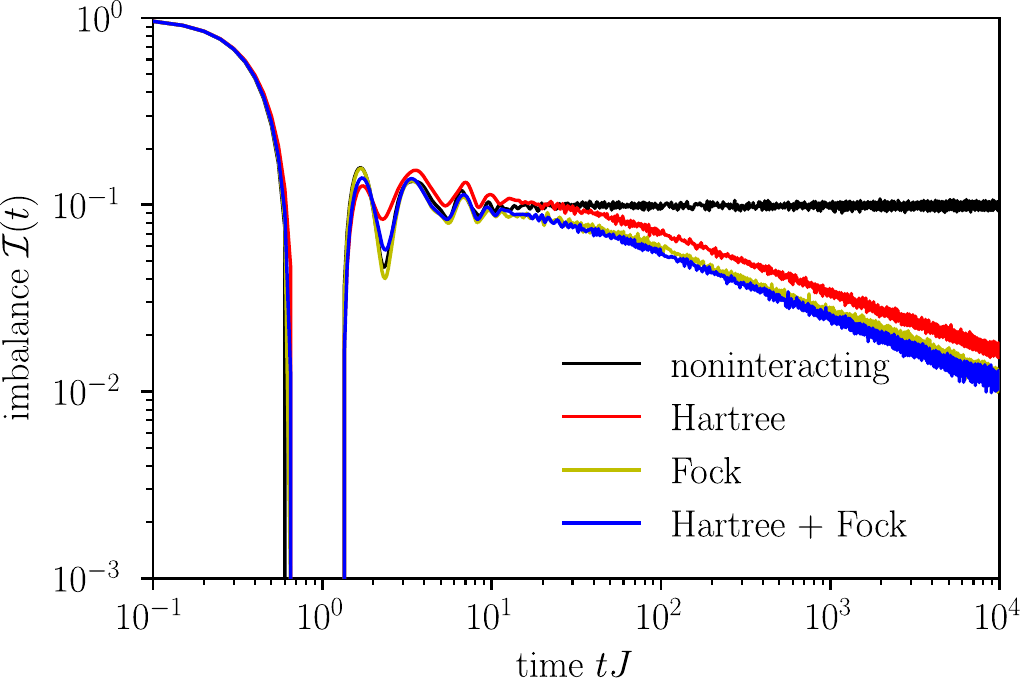}
	\caption{\textbf{Comparison between Hartree, Fock, and Hartree-Fock time evolution}. The time evolution of only Hartree (red), only Fock (yellow), and both Hartree and Fock (blue) is qualitatively similar. The time evolution is shown for random potentials of strength $W=2.0$ in the subdiffusive phase and are compared to the localized, non-interacting system with the same disorder (black). }
	\label{fig:H+F+HF} 
\end{figure}
Just by looking at the decay of the imbalance or the density-density correlation of an initial state in Sec.~\ref{sec:relax}, it is unclear whether this decay is only due to dephasing effects between different disorder samples and different lattice sites or true delocalization. In principle, this scenario can already be discarded by our results for the local spectral function in Sec.~\ref{sec:spectral}, which we computed for a single disorder realization. To further substantiate that the decay of the correlation functions is due to particle transport, we look at the occupation number $n_j(t)$ of a single site and a single disorder realization. In a delocalizing system, the single-site occupation number will approach $1/2$ for late times, $n_j(t\rightarrow \infty)=1/2$. On the Hartree-Fock level $n_j(t)$ contains oscillations which will persist forever and delocalization corresponds to these oscillations being centered around $1/2$. Decay of oscillations can only be obtained in higher order in the interaction and is a true many-particle effect. In order to remove the oscillations, we compute the time averaged occupations, $\lim_{t\to\infty} \frac{1}{t} \int_0^t ds~n_j(s) = 1/2$, which smoothly approach $1/2$ at late times when the system delocalizes. That can be observed in Fig.~\ref{fig:singlesite}, where we show that the time-averaged occupation number on a single site approaches $1/2$ for a few different disorder samples at weak disorder. For strong disorder the occupation remains close to its initial value, Fig.~\ref{fig:singlesite} as expected for a many-body localized system.

\section{Analysis of the Hartree and Fock contributions \label{sec:hfself}}

In Sec.~\ref{sec:noise}, we have focused on analyzing the Hartree self-energy. In Fig.~\ref{fig:H+F+HF} we show, that the time  evolution, when taking into account only Hartree, only  Fock, or both Hartree and Fock contributions, are qualitatively similar. This is why it is sufficient to focus on the Hartree selfenergy, $\Sigma_{jj}^\mathrm{H}(t)$, when we analyzing the self-consistent noise.

\bibliography{MBL_arxiv_v6.bib}

\begin{thebibliography}{72}%
\makeatletter
\providecommand \@ifxundefined [1]{%
 \@ifx{#1\undefined}
}%
\providecommand \@ifnum [1]{%
 \ifnum #1\expandafter \@firstoftwo
 \else \expandafter \@secondoftwo
 \fi
}%
\providecommand \@ifx [1]{%
 \ifx #1\expandafter \@firstoftwo
 \else \expandafter \@secondoftwo
 \fi
}%
\providecommand \natexlab [1]{#1}%
\providecommand \enquote  [1]{``#1''}%
\providecommand \bibnamefont  [1]{#1}%
\providecommand \bibfnamefont [1]{#1}%
\providecommand \citenamefont [1]{#1}%
\providecommand \href@noop [0]{\@secondoftwo}%
\providecommand \href [0]{\begingroup \@sanitize@url \@href}%
\providecommand \@href[1]{\@@startlink{#1}\@@href}%
\providecommand \@@href[1]{\endgroup#1\@@endlink}%
\providecommand \@sanitize@url [0]{\catcode `\\12\catcode `\$12\catcode
  `\&12\catcode `\#12\catcode `\^12\catcode `\_12\catcode `\%12\relax}%
\providecommand \@@startlink[1]{}%
\providecommand \@@endlink[0]{}%
\providecommand \url  [0]{\begingroup\@sanitize@url \@url }%
\providecommand \@url [1]{\endgroup\@href {#1}{\urlprefix }}%
\providecommand \urlprefix  [0]{URL }%
\providecommand \Eprint [0]{\href }%
\providecommand \doibase [0]{http://dx.doi.org/}%
\providecommand \selectlanguage [0]{\@gobble}%
\providecommand \bibinfo  [0]{\@secondoftwo}%
\providecommand \bibfield  [0]{\@secondoftwo}%
\providecommand \translation [1]{[#1]}%
\providecommand \BibitemOpen [0]{}%
\providecommand \bibitemStop [0]{}%
\providecommand \bibitemNoStop [0]{.\EOS\space}%
\providecommand \EOS [0]{\spacefactor3000\relax}%
\providecommand \BibitemShut  [1]{\csname bibitem#1\endcsname}%
\let\auto@bib@innerbib\@empty
\bibitem [{\citenamefont {{Schreiber, Michael and Hodgman, Sean S. and Bordia,
  Pranjal and L{\"u}schen, Henrik P. and Fischer, Mark H. and Vosk, Ronen and
  Altman, Ehud and Schneider, Ulrich and Bloch,
  Immanuel}}(2015)}]{schreiber2015observation}%
  \BibitemOpen
  \bibfield  {author} {\bibinfo {author} {\bibnamefont {{Schreiber, Michael and
  Hodgman, Sean S. and Bordia, Pranjal and L{\"u}schen, Henrik P. and Fischer,
  Mark H. and Vosk, Ronen and Altman, Ehud and Schneider, Ulrich and Bloch,
  Immanuel}}},\ }\bibfield  {title} {\enquote {\bibinfo {title} {Observation of
  many-body localization of interacting fermions in a quasi-random optical
  lattice},}\ }\href@noop {} {\bibfield  {journal} {\bibinfo  {journal}
  {Science}\ ,\ \bibinfo {pages} {7432}} (\bibinfo {year} {2015})}\BibitemShut
  {NoStop}%
\bibitem [{\citenamefont {Kondov}\ \emph {et~al.}(2015)\citenamefont {Kondov},
  \citenamefont {McGehee}, \citenamefont {Xu},\ and\ \citenamefont
  {DeMarco}}]{kondov2015disorder}%
  \BibitemOpen
  \bibfield  {author} {\bibinfo {author} {\bibfnamefont {S.~S.}\ \bibnamefont
  {Kondov}}, \bibinfo {author} {\bibfnamefont {W.~R.}\ \bibnamefont {McGehee}},
  \bibinfo {author} {\bibfnamefont {W.}~\bibnamefont {Xu}}, \ and\ \bibinfo
  {author} {\bibfnamefont {B.}~\bibnamefont {DeMarco}},\ }\bibfield  {title}
  {\enquote {\bibinfo {title} {Disorder-induced localization in a strongly
  correlated atomic hubbard gas},}\ }\href@noop {} {\bibfield  {journal}
  {\bibinfo  {journal} {Physical Review Letters}\ }\textbf {\bibinfo {volume}
  {114}},\ \bibinfo {pages} {083002} (\bibinfo {year} {2015})}\BibitemShut
  {NoStop}%
\bibitem [{\citenamefont {Smith}\ \emph {et~al.}(2016)\citenamefont {Smith},
  \citenamefont {Lee}, \citenamefont {Richerme}, \citenamefont {Neyenhuis},
  \citenamefont {Hess}, \citenamefont {Hauke}, \citenamefont {Heyl},
  \citenamefont {Huse},\ and\ \citenamefont {Monroe}}]{smith2016many}%
  \BibitemOpen
  \bibfield  {author} {\bibinfo {author} {\bibfnamefont {Jacob}\ \bibnamefont
  {Smith}}, \bibinfo {author} {\bibfnamefont {Aaron}\ \bibnamefont {Lee}},
  \bibinfo {author} {\bibfnamefont {Philip}\ \bibnamefont {Richerme}}, \bibinfo
  {author} {\bibfnamefont {Brian}\ \bibnamefont {Neyenhuis}}, \bibinfo {author}
  {\bibfnamefont {Paul~W.}\ \bibnamefont {Hess}}, \bibinfo {author}
  {\bibfnamefont {Philipp}\ \bibnamefont {Hauke}}, \bibinfo {author}
  {\bibfnamefont {Markus}\ \bibnamefont {Heyl}}, \bibinfo {author}
  {\bibfnamefont {David~A.}\ \bibnamefont {Huse}}, \ and\ \bibinfo {author}
  {\bibfnamefont {Christopher}\ \bibnamefont {Monroe}},\ }\bibfield  {title}
  {\enquote {\bibinfo {title} {Many-body localization in a quantum simulator
  with programmable random disorder},}\ }\href@noop {} {\bibfield  {journal}
  {\bibinfo  {journal} {Nature Physics}\ }\textbf {\bibinfo {volume} {12}},\
  \bibinfo {pages} {907} (\bibinfo {year} {2016})}\BibitemShut {NoStop}%
\bibitem [{\citenamefont {Bordia}\ \emph {et~al.}(2016)\citenamefont {Bordia},
  \citenamefont {L{\"u}schen}, \citenamefont {Hodgman}, \citenamefont
  {Schreiber}, \citenamefont {Bloch},\ and\ \citenamefont
  {Schneider}}]{bordia2016coupling}%
  \BibitemOpen
  \bibfield  {author} {\bibinfo {author} {\bibfnamefont {Pranjal}\ \bibnamefont
  {Bordia}}, \bibinfo {author} {\bibfnamefont {Henrik~P.}\ \bibnamefont
  {L{\"u}schen}}, \bibinfo {author} {\bibfnamefont {Sean~S.}\ \bibnamefont
  {Hodgman}}, \bibinfo {author} {\bibfnamefont {Michael}\ \bibnamefont
  {Schreiber}}, \bibinfo {author} {\bibfnamefont {Immanuel}\ \bibnamefont
  {Bloch}}, \ and\ \bibinfo {author} {\bibfnamefont {Ulrich}\ \bibnamefont
  {Schneider}},\ }\bibfield  {title} {\enquote {\bibinfo {title} {Coupling
  identical one-dimensional many-body localized systems},}\ }\href@noop {}
  {\bibfield  {journal} {\bibinfo  {journal} {Physical Review Letters}\
  }\textbf {\bibinfo {volume} {116}},\ \bibinfo {pages} {140401} (\bibinfo
  {year} {2016})}\BibitemShut {NoStop}%
\bibitem [{\citenamefont {Choi}\ \emph {et~al.}(2016)\citenamefont {Choi},
  \citenamefont {Hild}, \citenamefont {Zeiher}, \citenamefont {Schau{\ss}},
  \citenamefont {Rubio-Abadal}, \citenamefont {Yefsah}, \citenamefont
  {Khemani}, \citenamefont {Huse}, \citenamefont {Bloch},\ and\ \citenamefont
  {Gross}}]{choi2016exploring}%
  \BibitemOpen
  \bibfield  {author} {\bibinfo {author} {\bibfnamefont {Jae-yoon}\
  \bibnamefont {Choi}}, \bibinfo {author} {\bibfnamefont {Sebastian}\
  \bibnamefont {Hild}}, \bibinfo {author} {\bibfnamefont {Johannes}\
  \bibnamefont {Zeiher}}, \bibinfo {author} {\bibfnamefont {Peter}\
  \bibnamefont {Schau{\ss}}}, \bibinfo {author} {\bibfnamefont {Antonio}\
  \bibnamefont {Rubio-Abadal}}, \bibinfo {author} {\bibfnamefont {Tarik}\
  \bibnamefont {Yefsah}}, \bibinfo {author} {\bibfnamefont {Vedika}\
  \bibnamefont {Khemani}}, \bibinfo {author} {\bibfnamefont {David~A.}\
  \bibnamefont {Huse}}, \bibinfo {author} {\bibfnamefont {Immanuel}\
  \bibnamefont {Bloch}}, \ and\ \bibinfo {author} {\bibfnamefont {Christian}\
  \bibnamefont {Gross}},\ }\bibfield  {title} {\enquote {\bibinfo {title}
  {Exploring the many-body localization transition in two dimensions},}\
  }\href@noop {} {\bibfield  {journal} {\bibinfo  {journal} {Science}\ }\textbf
  {\bibinfo {volume} {352}},\ \bibinfo {pages} {1547--1552} (\bibinfo {year}
  {2016})}\BibitemShut {NoStop}%
\bibitem [{\citenamefont {Bordia}\ \emph
  {et~al.}(2017{\natexlab{a}})\citenamefont {Bordia}, \citenamefont
  {L{\"u}schen}, \citenamefont {Schneider}, \citenamefont {Knap},\ and\
  \citenamefont {Bloch}}]{bordia2017periodically}%
  \BibitemOpen
  \bibfield  {author} {\bibinfo {author} {\bibfnamefont {Pranjal}\ \bibnamefont
  {Bordia}}, \bibinfo {author} {\bibfnamefont {Henrik}\ \bibnamefont
  {L{\"u}schen}}, \bibinfo {author} {\bibfnamefont {Ulrich}\ \bibnamefont
  {Schneider}}, \bibinfo {author} {\bibfnamefont {Michael}\ \bibnamefont
  {Knap}}, \ and\ \bibinfo {author} {\bibfnamefont {Immanuel}\ \bibnamefont
  {Bloch}},\ }\bibfield  {title} {\enquote {\bibinfo {title} {Periodically
  driving a many-body localized quantum system},}\ }\href@noop {} {\bibfield
  {journal} {\bibinfo  {journal} {Nature Physics}\ }\textbf {\bibinfo {volume}
  {13}},\ \bibinfo {pages} {460} (\bibinfo {year}
  {2017}{\natexlab{a}})}\BibitemShut {NoStop}%
\bibitem [{\citenamefont {Bordia}\ \emph
  {et~al.}(2017{\natexlab{b}})\citenamefont {Bordia}, \citenamefont
  {L\"uschen}, \citenamefont {Scherg}, \citenamefont {Gopalakrishnan},
  \citenamefont {Knap}, \citenamefont {Schneider},\ and\ \citenamefont
  {Bloch}}]{bordia20172D}%
  \BibitemOpen
  \bibfield  {author} {\bibinfo {author} {\bibfnamefont {Pranjal}\ \bibnamefont
  {Bordia}}, \bibinfo {author} {\bibfnamefont {Henrik}\ \bibnamefont
  {L\"uschen}}, \bibinfo {author} {\bibfnamefont {Sebastian}\ \bibnamefont
  {Scherg}}, \bibinfo {author} {\bibfnamefont {Sarang}\ \bibnamefont
  {Gopalakrishnan}}, \bibinfo {author} {\bibfnamefont {Michael}\ \bibnamefont
  {Knap}}, \bibinfo {author} {\bibfnamefont {Ulrich}\ \bibnamefont
  {Schneider}}, \ and\ \bibinfo {author} {\bibfnamefont {Immanuel}\
  \bibnamefont {Bloch}},\ }\bibfield  {title} {\enquote {\bibinfo {title}
  {Probing slow relaxation and many-body localization in two-dimensional
  quasiperiodic systems},}\ }\href@noop {} {\bibfield  {journal} {\bibinfo
  {journal} {Phys. Rev. X}\ }\textbf {\bibinfo {volume} {7}},\ \bibinfo {pages}
  {041047} (\bibinfo {year} {2017}{\natexlab{b}})}\BibitemShut {NoStop}%
\bibitem [{\citenamefont {L{\"u}schen}\ \emph {et~al.}(2017)\citenamefont
  {L{\"u}schen}, \citenamefont {Bordia}, \citenamefont {Scherg}, \citenamefont
  {Alet}, \citenamefont {Altman}, \citenamefont {Schneider},\ and\
  \citenamefont {Bloch}}]{luschen2017observation}%
  \BibitemOpen
  \bibfield  {author} {\bibinfo {author} {\bibfnamefont {Henrik~P.}\
  \bibnamefont {L{\"u}schen}}, \bibinfo {author} {\bibfnamefont {Pranjal}\
  \bibnamefont {Bordia}}, \bibinfo {author} {\bibfnamefont {Sebastian}\
  \bibnamefont {Scherg}}, \bibinfo {author} {\bibfnamefont {Fabien}\
  \bibnamefont {Alet}}, \bibinfo {author} {\bibfnamefont {Ehud}\ \bibnamefont
  {Altman}}, \bibinfo {author} {\bibfnamefont {Ulrich}\ \bibnamefont
  {Schneider}}, \ and\ \bibinfo {author} {\bibfnamefont {Immanuel}\
  \bibnamefont {Bloch}},\ }\bibfield  {title} {\enquote {\bibinfo {title}
  {Observation of slow dynamics near the many-body localization transition in
  one-dimensional quasiperiodic systems},}\ }\href@noop {} {\bibfield
  {journal} {\bibinfo  {journal} {Physical Review Letters}\ }\textbf {\bibinfo
  {volume} {119}},\ \bibinfo {pages} {260401} (\bibinfo {year}
  {2017})}\BibitemShut {NoStop}%
\bibitem [{\citenamefont {Roushan}\ \emph {et~al.}(2017)\citenamefont
  {Roushan}, \citenamefont {Neill}, \citenamefont {Tangpanitanon},
  \citenamefont {Bastidas}, \citenamefont {Megrant}, \citenamefont {Barends},
  \citenamefont {Chen}, \citenamefont {Chen}, \citenamefont {Chiaro},
  \citenamefont {Dunsworth}, \citenamefont {Fowler}, \citenamefont {Foxen},
  \citenamefont {Giustina}, \citenamefont {Jeffrey}, \citenamefont {Kelly},
  \citenamefont {Lucero}, \citenamefont {Mutus}, \citenamefont {Neeley},
  \citenamefont {Quintana}, \citenamefont {Sank}, \citenamefont {Vainsencher},
  \citenamefont {Wenner}, \citenamefont {White}, \citenamefont {Neven},
  \citenamefont {Angelakis},\ and\ \citenamefont
  {Martinis}}]{roushanGoogle2017}%
  \BibitemOpen
  \bibfield  {author} {\bibinfo {author} {\bibfnamefont {P.}~\bibnamefont
  {Roushan}}, \bibinfo {author} {\bibfnamefont {C.}~\bibnamefont {Neill}},
  \bibinfo {author} {\bibfnamefont {J.}~\bibnamefont {Tangpanitanon}}, \bibinfo
  {author} {\bibfnamefont {V.~M.}\ \bibnamefont {Bastidas}}, \bibinfo {author}
  {\bibfnamefont {A.}~\bibnamefont {Megrant}}, \bibinfo {author} {\bibfnamefont
  {R.}~\bibnamefont {Barends}}, \bibinfo {author} {\bibfnamefont
  {Y.}~\bibnamefont {Chen}}, \bibinfo {author} {\bibfnamefont {Z.}~\bibnamefont
  {Chen}}, \bibinfo {author} {\bibfnamefont {B.}~\bibnamefont {Chiaro}},
  \bibinfo {author} {\bibfnamefont {A.}~\bibnamefont {Dunsworth}}, \bibinfo
  {author} {\bibfnamefont {A.}~\bibnamefont {Fowler}}, \bibinfo {author}
  {\bibfnamefont {B.}~\bibnamefont {Foxen}}, \bibinfo {author} {\bibfnamefont
  {M.}~\bibnamefont {Giustina}}, \bibinfo {author} {\bibfnamefont
  {E.}~\bibnamefont {Jeffrey}}, \bibinfo {author} {\bibfnamefont
  {J.}~\bibnamefont {Kelly}}, \bibinfo {author} {\bibfnamefont
  {E.}~\bibnamefont {Lucero}}, \bibinfo {author} {\bibfnamefont
  {J.}~\bibnamefont {Mutus}}, \bibinfo {author} {\bibfnamefont
  {M.}~\bibnamefont {Neeley}}, \bibinfo {author} {\bibfnamefont
  {C.}~\bibnamefont {Quintana}}, \bibinfo {author} {\bibfnamefont
  {D.}~\bibnamefont {Sank}}, \bibinfo {author} {\bibfnamefont {A.}~\bibnamefont
  {Vainsencher}}, \bibinfo {author} {\bibfnamefont {J.}~\bibnamefont {Wenner}},
  \bibinfo {author} {\bibfnamefont {T.}~\bibnamefont {White}}, \bibinfo
  {author} {\bibfnamefont {H.}~\bibnamefont {Neven}}, \bibinfo {author}
  {\bibfnamefont {D.~G.}\ \bibnamefont {Angelakis}}, \ and\ \bibinfo {author}
  {\bibfnamefont {J.}~\bibnamefont {Martinis}},\ }\bibfield  {title} {\enquote
  {\bibinfo {title} {Spectroscopic signatures of localization with interacting
  photons in superconducting qubits},}\ }\href@noop {} {\bibfield  {journal}
  {\bibinfo  {journal} {Science}\ }\textbf {\bibinfo {volume} {358}},\ \bibinfo
  {pages} {1175--1179} (\bibinfo {year} {2017})}\BibitemShut {NoStop}%
\bibitem [{\citenamefont {{Silevitch}}\ \emph {et~al.}(2017)\citenamefont
  {{Silevitch}}, \citenamefont {{Aeppli}},\ and\ \citenamefont
  {{Rosenbaum}}}]{rosenbaum2017}%
  \BibitemOpen
  \bibfield  {author} {\bibinfo {author} {\bibfnamefont {D.~M.}\ \bibnamefont
  {{Silevitch}}}, \bibinfo {author} {\bibfnamefont {G.}~\bibnamefont
  {{Aeppli}}}, \ and\ \bibinfo {author} {\bibfnamefont {T.~F.}\ \bibnamefont
  {{Rosenbaum}}},\ }\bibfield  {title} {\enquote {\bibinfo {title} {{Probing
  many-body localization in a disordered quantum magnet}},}\ }\href@noop {}
  {\bibfield  {journal} {\bibinfo  {journal} {arXiv:1707.04952}\ } (\bibinfo
  {year} {2017})}\BibitemShut {NoStop}%
\bibitem [{\citenamefont {Lukin}\ \emph {et~al.}(2018)\citenamefont {Lukin},
  \citenamefont {Rispoli}, \citenamefont {Schittko}, \citenamefont {Tai},
  \citenamefont {Kaufman}, \citenamefont {Choi}, \citenamefont {Khemani},
  \citenamefont {L{'e}onard},\ and\ \citenamefont
  {Greiner}}]{lukin_probing_2018}%
  \BibitemOpen
  \bibfield  {author} {\bibinfo {author} {\bibfnamefont {Alexander}\
  \bibnamefont {Lukin}}, \bibinfo {author} {\bibfnamefont {Matthew}\
  \bibnamefont {Rispoli}}, \bibinfo {author} {\bibfnamefont {Robert}\
  \bibnamefont {Schittko}}, \bibinfo {author} {\bibfnamefont {M.~Eric}\
  \bibnamefont {Tai}}, \bibinfo {author} {\bibfnamefont {Adam~M.}\ \bibnamefont
  {Kaufman}}, \bibinfo {author} {\bibfnamefont {Soonwon}\ \bibnamefont {Choi}},
  \bibinfo {author} {\bibfnamefont {Vedika}\ \bibnamefont {Khemani}}, \bibinfo
  {author} {\bibfnamefont {Julian}\ \bibnamefont {L{'e}onard}}, \ and\ \bibinfo
  {author} {\bibfnamefont {Markus}\ \bibnamefont {Greiner}},\ }\bibfield
  {title} {\enquote {\bibinfo {title} {Probing entanglement in a
  many-body-localized system},}\ }\href@noop {} {\bibfield  {journal} {\bibinfo
   {journal} {{arXiv}:1805.09819}\ } (\bibinfo {year} {2018})}\BibitemShut
  {NoStop}%
\bibitem [{\citenamefont {Basko}\ \emph {et~al.}(2006)\citenamefont {Basko},
  \citenamefont {Aleiner},\ and\ \citenamefont {Altshuler}}]{basko2006metal}%
  \BibitemOpen
  \bibfield  {author} {\bibinfo {author} {\bibfnamefont {D.~M.}\ \bibnamefont
  {Basko}}, \bibinfo {author} {\bibfnamefont {I.~L.}\ \bibnamefont {Aleiner}},
  \ and\ \bibinfo {author} {\bibfnamefont {B.~L.}\ \bibnamefont {Altshuler}},\
  }\bibfield  {title} {\enquote {\bibinfo {title} {Metal--insulator transition
  in a weakly interacting many-electron system with localized single-particle
  states},}\ }\href@noop {} {\bibfield  {journal} {\bibinfo  {journal} {Annals
  of physics}\ }\textbf {\bibinfo {volume} {321}},\ \bibinfo {pages}
  {1126--1205} (\bibinfo {year} {2006})}\BibitemShut {NoStop}%
\bibitem [{\citenamefont {Gornyi}\ \emph {et~al.}(2005)\citenamefont {Gornyi},
  \citenamefont {Mirlin},\ and\ \citenamefont
  {Polyakov}}]{gornyi2005interacting}%
  \BibitemOpen
  \bibfield  {author} {\bibinfo {author} {\bibfnamefont {I.~V.}\ \bibnamefont
  {Gornyi}}, \bibinfo {author} {\bibfnamefont {A.~D.}\ \bibnamefont {Mirlin}},
  \ and\ \bibinfo {author} {\bibfnamefont {D.~G.}\ \bibnamefont {Polyakov}},\
  }\bibfield  {title} {\enquote {\bibinfo {title} {Interacting electrons in
  disordered wires: Anderson localization and low-t transport},}\ }\href@noop
  {} {\bibfield  {journal} {\bibinfo  {journal} {Physical Review Letters}\
  }\textbf {\bibinfo {volume} {95}},\ \bibinfo {pages} {206603} (\bibinfo
  {year} {2005})}\BibitemShut {NoStop}%
\bibitem [{\citenamefont {Imbrie}(2016)}]{imbrie2016many}%
  \BibitemOpen
  \bibfield  {author} {\bibinfo {author} {\bibfnamefont {John~Z}\ \bibnamefont
  {Imbrie}},\ }\bibfield  {title} {\enquote {\bibinfo {title} {On many-body
  localization for quantum spin chains},}\ }\href@noop {} {\bibfield  {journal}
  {\bibinfo  {journal} {Journal of Statistical Physics}\ }\textbf {\bibinfo
  {volume} {163}},\ \bibinfo {pages} {998--1048} (\bibinfo {year}
  {2016})}\BibitemShut {NoStop}%
\bibitem [{\citenamefont {Nandkishore}\ and\ \citenamefont
  {Huse}(2015)}]{nandkishore2015many}%
  \BibitemOpen
  \bibfield  {author} {\bibinfo {author} {\bibfnamefont {Rahul}\ \bibnamefont
  {Nandkishore}}\ and\ \bibinfo {author} {\bibfnamefont {David~A.}\
  \bibnamefont {Huse}},\ }\bibfield  {title} {\enquote {\bibinfo {title}
  {Many-body localization and thermalization in quantum statistical
  mechanics},}\ }\href@noop {} {\bibfield  {journal} {\bibinfo  {journal}
  {Annu. Rev. Condens. Matter Phys.}\ }\textbf {\bibinfo {volume} {6}},\
  \bibinfo {pages} {15--38} (\bibinfo {year} {2015})}\BibitemShut {NoStop}%
\bibitem [{\citenamefont {Altman}\ and\ \citenamefont
  {Vosk}(2015)}]{altman2015universal}%
  \BibitemOpen
  \bibfield  {author} {\bibinfo {author} {\bibfnamefont {Ehud}\ \bibnamefont
  {Altman}}\ and\ \bibinfo {author} {\bibfnamefont {Ronen}\ \bibnamefont
  {Vosk}},\ }\bibfield  {title} {\enquote {\bibinfo {title} {Universal dynamics
  and renormalization in many-body-localized systems},}\ }\href@noop {}
  {\bibfield  {journal} {\bibinfo  {journal} {Annu. Rev. Condens. Matter
  Phys.}\ }\textbf {\bibinfo {volume} {6}},\ \bibinfo {pages} {383--409}
  (\bibinfo {year} {2015})}\BibitemShut {NoStop}%
\bibitem [{\citenamefont {Deutsch}(1991)}]{deutsch1991quantum}%
  \BibitemOpen
  \bibfield  {author} {\bibinfo {author} {\bibfnamefont {Josh~M}\ \bibnamefont
  {Deutsch}},\ }\bibfield  {title} {\enquote {\bibinfo {title} {Quantum
  statistical mechanics in a closed system},}\ }\href@noop {} {\bibfield
  {journal} {\bibinfo  {journal} {Physical Review A}\ }\textbf {\bibinfo
  {volume} {43}},\ \bibinfo {pages} {2046} (\bibinfo {year}
  {1991})}\BibitemShut {NoStop}%
\bibitem [{\citenamefont {Srednicki}(1994)}]{srednicki1994chaos}%
  \BibitemOpen
  \bibfield  {author} {\bibinfo {author} {\bibfnamefont {Mark}\ \bibnamefont
  {Srednicki}},\ }\bibfield  {title} {\enquote {\bibinfo {title} {Chaos and
  quantum thermalization},}\ }\href@noop {} {\bibfield  {journal} {\bibinfo
  {journal} {Physical Review E}\ }\textbf {\bibinfo {volume} {50}},\ \bibinfo
  {pages} {888} (\bibinfo {year} {1994})}\BibitemShut {NoStop}%
\bibitem [{\citenamefont {Rigol}\ \emph {et~al.}(2008)\citenamefont {Rigol},
  \citenamefont {Dunjko},\ and\ \citenamefont
  {Olshanii}}]{rigol2008thermalization}%
  \BibitemOpen
  \bibfield  {author} {\bibinfo {author} {\bibfnamefont {Marcos}\ \bibnamefont
  {Rigol}}, \bibinfo {author} {\bibfnamefont {Vanja}\ \bibnamefont {Dunjko}}, \
  and\ \bibinfo {author} {\bibfnamefont {Maxim}\ \bibnamefont {Olshanii}},\
  }\bibfield  {title} {\enquote {\bibinfo {title} {Thermalization and its
  mechanism for generic isolated quantum systems},}\ }\href@noop {} {\bibfield
  {journal} {\bibinfo  {journal} {Nature}\ }\textbf {\bibinfo {volume} {452}},\
  \bibinfo {pages} {854} (\bibinfo {year} {2008})}\BibitemShut {NoStop}%
\bibitem [{\citenamefont {De~Chiara}\ \emph {et~al.}(2006)\citenamefont
  {De~Chiara}, \citenamefont {Montangero}, \citenamefont {Calabrese},\ and\
  \citenamefont {Fazio}}]{de2006entanglement}%
  \BibitemOpen
  \bibfield  {author} {\bibinfo {author} {\bibfnamefont {Gabriele}\
  \bibnamefont {De~Chiara}}, \bibinfo {author} {\bibfnamefont {Simone}\
  \bibnamefont {Montangero}}, \bibinfo {author} {\bibfnamefont {Pasquale}\
  \bibnamefont {Calabrese}}, \ and\ \bibinfo {author} {\bibfnamefont {Rosario}\
  \bibnamefont {Fazio}},\ }\bibfield  {title} {\enquote {\bibinfo {title}
  {Entanglement entropy dynamics of heisenberg chains},}\ }\href@noop {}
  {\bibfield  {journal} {\bibinfo  {journal} {Journal of Statistical Mechanics:
  Theory and Experiment}\ }\textbf {\bibinfo {volume} {2006}},\ \bibinfo
  {pages} {P03001} (\bibinfo {year} {2006})}\BibitemShut {NoStop}%
\bibitem [{\citenamefont {\ifmmode \check{Z}\else
  \v{Z}\fi{}nidari\ifmmode~\check{c}\else \v{c}\fi{}}\ \emph
  {et~al.}(2008)\citenamefont {\ifmmode \check{Z}\else
  \v{Z}\fi{}nidari\ifmmode~\check{c}\else \v{c}\fi{}}, \citenamefont {Prosen},\
  and\ \citenamefont {Prelov\ifmmode~\check{s}\else
  \v{s}\fi{}ek}}]{ZnidaricPrelovsek08}%
  \BibitemOpen
  \bibfield  {author} {\bibinfo {author} {\bibfnamefont {Marko}\ \bibnamefont
  {\ifmmode \check{Z}\else \v{Z}\fi{}nidari\ifmmode~\check{c}\else
  \v{c}\fi{}}}, \bibinfo {author} {\bibfnamefont
  {Toma\ifmmode\check{z}\else\v{z}\fi{}}\ \bibnamefont {Prosen}}, \ and\
  \bibinfo {author} {\bibfnamefont {Peter}\ \bibnamefont
  {Prelov\ifmmode~\check{s}\else \v{s}\fi{}ek}},\ }\bibfield  {title} {\enquote
  {\bibinfo {title} {Many-body localization in the {H}eisenberg {XXZ} magnet in
  a random field},}\ }\href@noop {} {\bibfield  {journal} {\bibinfo  {journal}
  {Phys. Rev. B}\ }\textbf {\bibinfo {volume} {77}},\ \bibinfo {pages} {064426}
  (\bibinfo {year} {2008})}\BibitemShut {NoStop}%
\bibitem [{\citenamefont {Bardarson}\ \emph {et~al.}(2012)\citenamefont
  {Bardarson}, \citenamefont {Pollmann},\ and\ \citenamefont
  {Moore}}]{bardarson2012unbounded}%
  \BibitemOpen
  \bibfield  {author} {\bibinfo {author} {\bibfnamefont {Jens~H}\ \bibnamefont
  {Bardarson}}, \bibinfo {author} {\bibfnamefont {Frank}\ \bibnamefont
  {Pollmann}}, \ and\ \bibinfo {author} {\bibfnamefont {Joel~E}\ \bibnamefont
  {Moore}},\ }\bibfield  {title} {\enquote {\bibinfo {title} {Unbounded growth
  of entanglement in models of many-body localization},}\ }\href@noop {}
  {\bibfield  {journal} {\bibinfo  {journal} {Physical Review Letters}\
  }\textbf {\bibinfo {volume} {109}},\ \bibinfo {pages} {017202} (\bibinfo
  {year} {2012})}\BibitemShut {NoStop}%
\bibitem [{\citenamefont {Serbyn}\ \emph
  {et~al.}(2014{\natexlab{a}})\citenamefont {Serbyn}, \citenamefont {Knap},
  \citenamefont {Gopalakrishnan}, \citenamefont {Papi{\'c}}, \citenamefont
  {Yao}, \citenamefont {Laumann}, \citenamefont {Abanin}, \citenamefont
  {Lukin},\ and\ \citenamefont {Demler}}]{serbyn2014interferometric}%
  \BibitemOpen
  \bibfield  {author} {\bibinfo {author} {\bibfnamefont {M.}~\bibnamefont
  {Serbyn}}, \bibinfo {author} {\bibfnamefont {Michael}\ \bibnamefont {Knap}},
  \bibinfo {author} {\bibfnamefont {Sarang}\ \bibnamefont {Gopalakrishnan}},
  \bibinfo {author} {\bibfnamefont {Z.}~\bibnamefont {Papi{\'c}}}, \bibinfo
  {author} {\bibfnamefont {Norman~Ying}\ \bibnamefont {Yao}}, \bibinfo {author}
  {\bibfnamefont {C.~R.}\ \bibnamefont {Laumann}}, \bibinfo {author}
  {\bibfnamefont {D.~A.}\ \bibnamefont {Abanin}}, \bibinfo {author}
  {\bibfnamefont {Mikhail~D.}\ \bibnamefont {Lukin}}, \ and\ \bibinfo {author}
  {\bibfnamefont {Eugene~A.}\ \bibnamefont {Demler}},\ }\bibfield  {title}
  {\enquote {\bibinfo {title} {Interferometric probes of many-body
  localization},}\ }\href@noop {} {\bibfield  {journal} {\bibinfo  {journal}
  {Physical Review Letters}\ }\textbf {\bibinfo {volume} {113}},\ \bibinfo
  {pages} {147204} (\bibinfo {year} {2014}{\natexlab{a}})}\BibitemShut
  {NoStop}%
\bibitem [{\citenamefont {Bahri}\ \emph {et~al.}(2015)\citenamefont {Bahri},
  \citenamefont {Vosk}, \citenamefont {Altman},\ and\ \citenamefont
  {Vishwanath}}]{bahri2015localization}%
  \BibitemOpen
  \bibfield  {author} {\bibinfo {author} {\bibfnamefont {Yasaman}\ \bibnamefont
  {Bahri}}, \bibinfo {author} {\bibfnamefont {Ronen}\ \bibnamefont {Vosk}},
  \bibinfo {author} {\bibfnamefont {Ehud}\ \bibnamefont {Altman}}, \ and\
  \bibinfo {author} {\bibfnamefont {Ashvin}\ \bibnamefont {Vishwanath}},\
  }\bibfield  {title} {\enquote {\bibinfo {title} {Localization and topology
  protected quantum coherence at the edge of hot matter},}\ }\href@noop {}
  {\bibfield  {journal} {\bibinfo  {journal} {Nature communications}\ }\textbf
  {\bibinfo {volume} {6}},\ \bibinfo {pages} {7341} (\bibinfo {year}
  {2015})}\BibitemShut {NoStop}%
\bibitem [{\citenamefont {Serbyn}\ \emph
  {et~al.}(2014{\natexlab{b}})\citenamefont {Serbyn}, \citenamefont
  {Papi\ifmmode~\acute{c}\else \'{c}\fi{}},\ and\ \citenamefont
  {Abanin}}]{serbynPapicAbaninQuantumQuenchesPRB201567}%
  \BibitemOpen
  \bibfield  {author} {\bibinfo {author} {\bibfnamefont {Maksym}\ \bibnamefont
  {Serbyn}}, \bibinfo {author} {\bibfnamefont {Z.}~\bibnamefont
  {Papi\ifmmode~\acute{c}\else \'{c}\fi{}}}, \ and\ \bibinfo {author}
  {\bibfnamefont {D.~A.}\ \bibnamefont {Abanin}},\ }\bibfield  {title}
  {\enquote {\bibinfo {title} {Quantum quenches in the many-body localized
  phase},}\ }\href@noop {} {\bibfield  {journal} {\bibinfo  {journal} {Phys.
  Rev. B}\ }\textbf {\bibinfo {volume} {90}},\ \bibinfo {pages} {174302}
  (\bibinfo {year} {2014}{\natexlab{b}})}\BibitemShut {NoStop}%
\bibitem [{\citenamefont {Torres-Herrera}\ \emph {et~al.}(2018)\citenamefont
  {Torres-Herrera}, \citenamefont {Garcia-Garcia},\ and\ \citenamefont
  {Santos}}]{torres-herrera_generic_2018}%
  \BibitemOpen
  \bibfield  {author} {\bibinfo {author} {\bibfnamefont {E.~J.}\ \bibnamefont
  {Torres-Herrera}}, \bibinfo {author} {\bibfnamefont {Antonio~M.}\
  \bibnamefont {Garcia-Garcia}}, \ and\ \bibinfo {author} {\bibfnamefont
  {Lea~F.}\ \bibnamefont {Santos}},\ }\bibfield  {title} {\enquote {\bibinfo
  {title} {Generic dynamical features of quenched interacting quantum systems:
  Survival probability, density imbalance, and out-of-time-ordered
  correlator},}\ }\href@noop {} {\bibfield  {journal} {\bibinfo  {journal}
  {Phys. Rev. B}\ }\textbf {\bibinfo {volume} {97}},\ \bibinfo {pages} {060303}
  (\bibinfo {year} {2018})}\BibitemShut {NoStop}%
\bibitem [{\citenamefont {Schiulaz}\ \emph {et~al.}(2018)\citenamefont
  {Schiulaz}, \citenamefont {Torres-Herrera},\ and\ \citenamefont
  {Santos}}]{schiulaz_thouless_2018}%
  \BibitemOpen
  \bibfield  {author} {\bibinfo {author} {\bibfnamefont {Mauro}\ \bibnamefont
  {Schiulaz}}, \bibinfo {author} {\bibfnamefont {E.~Jonathan}\ \bibnamefont
  {Torres-Herrera}}, \ and\ \bibinfo {author} {\bibfnamefont {Lea~F.}\
  \bibnamefont {Santos}},\ }\bibfield  {title} {\enquote {\bibinfo {title}
  {Thouless and relaxation time scales in many-body quantum systems},}\
  }\href@noop {} {\bibfield  {journal} {\bibinfo  {journal} {{arXiv}:1807.07577
  [cond-mat]}\ } (\bibinfo {year} {2018})}\BibitemShut {NoStop}%
\bibitem [{\citenamefont {Pal}\ and\ \citenamefont {Huse}(2010)}]{pal}%
  \BibitemOpen
  \bibfield  {author} {\bibinfo {author} {\bibfnamefont {Arijeet}\ \bibnamefont
  {Pal}}\ and\ \bibinfo {author} {\bibfnamefont {David~A.}\ \bibnamefont
  {Huse}},\ }\bibfield  {title} {\enquote {\bibinfo {title} {Many-body
  localization phase transition},}\ }\href@noop {} {\bibfield  {journal}
  {\bibinfo  {journal} {Phys. Rev. B}\ }\textbf {\bibinfo {volume} {82}},\
  \bibinfo {pages} {174411} (\bibinfo {year} {2010})}\BibitemShut {NoStop}%
\bibitem [{\citenamefont {Luitz}\ \emph {et~al.}(2015)\citenamefont {Luitz},
  \citenamefont {Laflorencie},\ and\ \citenamefont {Alet}}]{luitz2015many}%
  \BibitemOpen
  \bibfield  {author} {\bibinfo {author} {\bibfnamefont {David~J}\ \bibnamefont
  {Luitz}}, \bibinfo {author} {\bibfnamefont {Nicolas}\ \bibnamefont
  {Laflorencie}}, \ and\ \bibinfo {author} {\bibfnamefont {Fabien}\
  \bibnamefont {Alet}},\ }\bibfield  {title} {\enquote {\bibinfo {title}
  {Many-body localization edge in the random-field heisenberg chain},}\
  }\href@noop {} {\bibfield  {journal} {\bibinfo  {journal} {Physical Review
  B}\ }\textbf {\bibinfo {volume} {91}},\ \bibinfo {pages} {081103} (\bibinfo
  {year} {2015})}\BibitemShut {NoStop}%
\bibitem [{\citenamefont {Agarwal}\ \emph {et~al.}(2015)\citenamefont
  {Agarwal}, \citenamefont {Gopalakrishnan}, \citenamefont {Knap},
  \citenamefont {M{\"u}ller},\ and\ \citenamefont
  {Demler}}]{agarwal2015anomalous}%
  \BibitemOpen
  \bibfield  {author} {\bibinfo {author} {\bibfnamefont {Kartiek}\ \bibnamefont
  {Agarwal}}, \bibinfo {author} {\bibfnamefont {Sarang}\ \bibnamefont
  {Gopalakrishnan}}, \bibinfo {author} {\bibfnamefont {Michael}\ \bibnamefont
  {Knap}}, \bibinfo {author} {\bibfnamefont {Markus}\ \bibnamefont
  {M{\"u}ller}}, \ and\ \bibinfo {author} {\bibfnamefont {Eugene}\ \bibnamefont
  {Demler}},\ }\bibfield  {title} {\enquote {\bibinfo {title} {Anomalous
  diffusion and griffiths effects near the many-body localization
  transition},}\ }\href@noop {} {\bibfield  {journal} {\bibinfo  {journal}
  {Physical Review Letters}\ }\textbf {\bibinfo {volume} {114}},\ \bibinfo
  {pages} {160401} (\bibinfo {year} {2015})}\BibitemShut {NoStop}%
\bibitem [{\citenamefont {Vosk}\ \emph {et~al.}(2015)\citenamefont {Vosk},
  \citenamefont {Huse},\ and\ \citenamefont {Altman}}]{voskhusealtman}%
  \BibitemOpen
  \bibfield  {author} {\bibinfo {author} {\bibfnamefont {Ronen}\ \bibnamefont
  {Vosk}}, \bibinfo {author} {\bibfnamefont {David~A.}\ \bibnamefont {Huse}}, \
  and\ \bibinfo {author} {\bibfnamefont {Ehud}\ \bibnamefont {Altman}},\
  }\bibfield  {title} {\enquote {\bibinfo {title} {Theory of the many-body
  localization transition in one-dimensional systems},}\ }\href@noop {}
  {\bibfield  {journal} {\bibinfo  {journal} {Phys. Rev. X}\ }\textbf {\bibinfo
  {volume} {5}},\ \bibinfo {pages} {031032} (\bibinfo {year}
  {2015})}\BibitemShut {NoStop}%
\bibitem [{\citenamefont {Potter}\ and\ \citenamefont
  {Vasseur}(2016)}]{pottervasseursid}%
  \BibitemOpen
  \bibfield  {author} {\bibinfo {author} {\bibfnamefont {Andrew~C.}\
  \bibnamefont {Potter}}\ and\ \bibinfo {author} {\bibfnamefont {Romain}\
  \bibnamefont {Vasseur}},\ }\bibfield  {title} {\enquote {\bibinfo {title}
  {Symmetry constraints on many-body localization},}\ }\href@noop {} {\bibfield
   {journal} {\bibinfo  {journal} {Phys. Rev. B}\ }\textbf {\bibinfo {volume}
  {94}},\ \bibinfo {pages} {224206} (\bibinfo {year} {2016})}\BibitemShut
  {NoStop}%
\bibitem [{\citenamefont {Gopalakrishnan}\ and\ \citenamefont
  {Nandkishore}(2014)}]{gopalakrishnan_mean-field_2014-1}%
  \BibitemOpen
  \bibfield  {author} {\bibinfo {author} {\bibfnamefont {Sarang}\ \bibnamefont
  {Gopalakrishnan}}\ and\ \bibinfo {author} {\bibfnamefont {Rahul}\
  \bibnamefont {Nandkishore}},\ }\bibfield  {title} {\enquote {\bibinfo {title}
  {Mean-field theory of nearly many-body localized metals},}\ }\href@noop {}
  {\bibfield  {journal} {\bibinfo  {journal} {Phys. Rev. B}\ }\textbf {\bibinfo
  {volume} {90}},\ \bibinfo {pages} {224203} (\bibinfo {year}
  {2014})}\BibitemShut {NoStop}%
\bibitem [{\citenamefont {Dumitrescu}\ \emph {et~al.}(2017)\citenamefont
  {Dumitrescu}, \citenamefont {Vasseur},\ and\ \citenamefont {Potter}}]{dvp}%
  \BibitemOpen
  \bibfield  {author} {\bibinfo {author} {\bibfnamefont {Philipp~T.}\
  \bibnamefont {Dumitrescu}}, \bibinfo {author} {\bibfnamefont {Romain}\
  \bibnamefont {Vasseur}}, \ and\ \bibinfo {author} {\bibfnamefont {Andrew~C.}\
  \bibnamefont {Potter}},\ }\bibfield  {title} {\enquote {\bibinfo {title}
  {Scaling theory of entanglement at the many-body localization transition},}\
  }\href@noop {} {\bibfield  {journal} {\bibinfo  {journal} {Phys. Rev. Lett.}\
  }\textbf {\bibinfo {volume} {119}},\ \bibinfo {pages} {110604} (\bibinfo
  {year} {2017})}\BibitemShut {NoStop}%
\bibitem [{\citenamefont {Thiery}\ \emph {et~al.}(2017)\citenamefont {Thiery},
  \citenamefont {Huveneers}, \citenamefont {M{\"u}ller},\ and\ \citenamefont
  {De~Roeck}}]{thiery2017}%
  \BibitemOpen
  \bibfield  {author} {\bibinfo {author} {\bibfnamefont {Thimoth{\'e}e}\
  \bibnamefont {Thiery}}, \bibinfo {author} {\bibfnamefont {Fran{\c{c}}ois}\
  \bibnamefont {Huveneers}}, \bibinfo {author} {\bibfnamefont {Markus}\
  \bibnamefont {M{\"u}ller}}, \ and\ \bibinfo {author} {\bibfnamefont
  {Wojciech}\ \bibnamefont {De~Roeck}},\ }\bibfield  {title} {\enquote
  {\bibinfo {title} {Many-body delocalization as a quantum avalanche},}\
  }\href@noop {} {\bibfield  {journal} {\bibinfo  {journal} {arXiv:1706.09338}\
  } (\bibinfo {year} {2017})}\BibitemShut {NoStop}%
\bibitem [{\citenamefont {Khemani}\ \emph {et~al.}(2017)\citenamefont
  {Khemani}, \citenamefont {Lim}, \citenamefont {Sheng},\ and\ \citenamefont
  {Huse}}]{khemani_critical_2017}%
  \BibitemOpen
  \bibfield  {author} {\bibinfo {author} {\bibfnamefont {Vedika}\ \bibnamefont
  {Khemani}}, \bibinfo {author} {\bibfnamefont {S. P.}\ \bibnamefont {Lim}},
  \bibinfo {author} {\bibfnamefont {D. N.}\ \bibnamefont {Sheng}}, \ and\
  \bibinfo {author} {\bibfnamefont {David~A.}\ \bibnamefont {Huse}},\
  }\bibfield  {title} {\enquote {\bibinfo {title} {Critical properties of the
  many-body localization transition},}\ }\href@noop {} {\bibfield  {journal}
  {\bibinfo  {journal} {Phys. Rev. X}\ }\textbf {\bibinfo {volume} {7}}
  (\bibinfo {year} {2017})}\BibitemShut {NoStop}%
\bibitem [{\citenamefont {Gornyi}\ \emph {et~al.}(2017)\citenamefont {Gornyi},
  \citenamefont {Mirlin}, \citenamefont {Polyakov},\ and\ \citenamefont
  {Burin}}]{gornyi_spectral_2017}%
  \BibitemOpen
  \bibfield  {author} {\bibinfo {author} {\bibfnamefont {I.~V.}\ \bibnamefont
  {Gornyi}}, \bibinfo {author} {\bibfnamefont {A.~D.}\ \bibnamefont {Mirlin}},
  \bibinfo {author} {\bibfnamefont {D.~G.}\ \bibnamefont {Polyakov}}, \ and\
  \bibinfo {author} {\bibfnamefont {A.~L.}\ \bibnamefont {Burin}},\ }\bibfield
  {title} {\enquote {\bibinfo {title} {Spectral diffusion and scaling of
  many-body delocalization transitions},}\ }\href@noop {} {\bibfield  {journal}
  {\bibinfo  {journal} {Annalen der Physik}\ }\textbf {\bibinfo {volume}
  {529}},\ \bibinfo {pages} {1600360} (\bibinfo {year} {2017})}\BibitemShut
  {NoStop}%
\bibitem [{\citenamefont {{Goremykina}}\ \emph {et~al.}(2018)\citenamefont
  {{Goremykina}}, \citenamefont {{Vasseur}},\ and\ \citenamefont
  {{Serbyn}}}]{goremykina18}%
  \BibitemOpen
  \bibfield  {author} {\bibinfo {author} {\bibfnamefont {Anna}\ \bibnamefont
  {{Goremykina}}}, \bibinfo {author} {\bibfnamefont {Romain}\ \bibnamefont
  {{Vasseur}}}, \ and\ \bibinfo {author} {\bibfnamefont {Maksym}\ \bibnamefont
  {{Serbyn}}},\ }\bibfield  {title} {\enquote {\bibinfo {title} {{Analytically
  solvable renormalization group for the many-body localization transition}},}\
  }\href@noop {} {\bibfield  {journal} {\bibinfo  {journal} {arXiv:1807.04285}\
  } (\bibinfo {year} {2018})}\BibitemShut {NoStop}%
\bibitem [{\citenamefont {Han}\ and\ \citenamefont
  {Kim}(2018)}]{han_boltzmann_2018}%
  \BibitemOpen
  \bibfield  {author} {\bibinfo {author} {\bibfnamefont {Jae-Ho}\ \bibnamefont
  {Han}}\ and\ \bibinfo {author} {\bibfnamefont {Ki-Seok}\ \bibnamefont
  {Kim}},\ }\bibfield  {title} {\enquote {\bibinfo {title} {Boltzmann transport
  theory for many-body localization},}\ }\href@noop {} {\bibfield  {journal}
  {\bibinfo  {journal} {Phys. Rev. B}\ }\textbf {\bibinfo {volume} {97}},\
  \bibinfo {pages} {214206} (\bibinfo {year} {2018})}\BibitemShut {NoStop}%
\bibitem [{\citenamefont {Chaikin}\ and\ \citenamefont
  {Lubensky}(1995)}]{chaikinlubensky}%
  \BibitemOpen
  \bibfield  {author} {\bibinfo {author} {\bibfnamefont {P.~M.}\ \bibnamefont
  {Chaikin}}\ and\ \bibinfo {author} {\bibfnamefont {T.~C.}\ \bibnamefont
  {Lubensky}},\ }\href@noop {} {\emph {\bibinfo {title} {Principles of
  Condensed Matter Physics}}}\ (\bibinfo  {publisher} {Cambridge University
  Press},\ \bibinfo {year} {1995})\BibitemShut {NoStop}%
\bibitem [{\citenamefont {Vosk}\ and\ \citenamefont
  {Altman}(2013)}]{vosk2013many}%
  \BibitemOpen
  \bibfield  {author} {\bibinfo {author} {\bibfnamefont {Ronen}\ \bibnamefont
  {Vosk}}\ and\ \bibinfo {author} {\bibfnamefont {Ehud}\ \bibnamefont
  {Altman}},\ }\bibfield  {title} {\enquote {\bibinfo {title} {Many-body
  localization in one dimension as a dynamical renormalization group fixed
  point},}\ }\href@noop {} {\bibfield  {journal} {\bibinfo  {journal} {Physical
  Review Letters}\ }\textbf {\bibinfo {volume} {110}},\ \bibinfo {pages}
  {067204} (\bibinfo {year} {2013})}\BibitemShut {NoStop}%
\bibitem [{\citenamefont {Huse}\ \emph {et~al.}(2014)\citenamefont {Huse},
  \citenamefont {Nandkishore},\ and\ \citenamefont
  {Oganesyan}}]{huse2014phenomenology}%
  \BibitemOpen
  \bibfield  {author} {\bibinfo {author} {\bibfnamefont {David~A.}\
  \bibnamefont {Huse}}, \bibinfo {author} {\bibfnamefont {Rahul}\ \bibnamefont
  {Nandkishore}}, \ and\ \bibinfo {author} {\bibfnamefont {Vadim}\ \bibnamefont
  {Oganesyan}},\ }\bibfield  {title} {\enquote {\bibinfo {title} {Phenomenology
  of fully many-body-localized systems},}\ }\href@noop {} {\bibfield  {journal}
  {\bibinfo  {journal} {Physical Review B}\ }\textbf {\bibinfo {volume} {90}},\
  \bibinfo {pages} {174202} (\bibinfo {year} {2014})}\BibitemShut {NoStop}%
\bibitem [{\citenamefont {Serbyn}\ \emph {et~al.}(2013)\citenamefont {Serbyn},
  \citenamefont {Papi{\'c}},\ and\ \citenamefont {Abanin}}]{serbyn2013local}%
  \BibitemOpen
  \bibfield  {author} {\bibinfo {author} {\bibfnamefont {Maksym}\ \bibnamefont
  {Serbyn}}, \bibinfo {author} {\bibfnamefont {Z.}~\bibnamefont {Papi{\'c}}}, \
  and\ \bibinfo {author} {\bibfnamefont {Dmitry~A.}\ \bibnamefont {Abanin}},\
  }\bibfield  {title} {\enquote {\bibinfo {title} {Local conservation laws and
  the structure of the many-body localized states},}\ }\href@noop {} {\bibfield
   {journal} {\bibinfo  {journal} {Physical Review Letters}\ }\textbf {\bibinfo
  {volume} {111}},\ \bibinfo {pages} {127201} (\bibinfo {year}
  {2013})}\BibitemShut {NoStop}%
\bibitem [{\citenamefont {Ros}\ \emph {et~al.}(2015)\citenamefont {Ros},
  \citenamefont {M{\"u}ller},\ and\ \citenamefont
  {Scardicchio}}]{ros2015integrals}%
  \BibitemOpen
  \bibfield  {author} {\bibinfo {author} {\bibfnamefont {V.}~\bibnamefont
  {Ros}}, \bibinfo {author} {\bibfnamefont {M.}~\bibnamefont {M{\"u}ller}}, \
  and\ \bibinfo {author} {\bibfnamefont {A.}~\bibnamefont {Scardicchio}},\
  }\bibfield  {title} {\enquote {\bibinfo {title} {Integrals of motion in the
  many-body localized phase},}\ }\href@noop {} {\bibfield  {journal} {\bibinfo
  {journal} {Nuclear Physics B}\ }\textbf {\bibinfo {volume} {891}},\ \bibinfo
  {pages} {420--465} (\bibinfo {year} {2015})}\BibitemShut {NoStop}%
\bibitem [{\citenamefont {Baym}\ and\ \citenamefont
  {Kadanoff}(1961)}]{baym1961conservation}%
  \BibitemOpen
  \bibfield  {author} {\bibinfo {author} {\bibfnamefont {Gordon}\ \bibnamefont
  {Baym}}\ and\ \bibinfo {author} {\bibfnamefont {Leo~P.}\ \bibnamefont
  {Kadanoff}},\ }\bibfield  {title} {\enquote {\bibinfo {title} {Conservation
  laws and correlation functions},}\ }\href@noop {} {\bibfield  {journal}
  {\bibinfo  {journal} {Physical Review}\ }\textbf {\bibinfo {volume} {124}},\
  \bibinfo {pages} {287} (\bibinfo {year} {1961})}\BibitemShut {NoStop}%
\bibitem [{\citenamefont {Keldysh}\ \emph {et~al.}(1965)\citenamefont {Keldysh}
  \emph {et~al.}}]{keldysh1965diagram}%
  \BibitemOpen
  \bibfield  {author} {\bibinfo {author} {\bibfnamefont {Leonid~V.}\
  \bibnamefont {Keldysh}} \emph {et~al.},\ }\bibfield  {title} {\enquote
  {\bibinfo {title} {Diagram technique for nonequilibrium processes},}\
  }\href@noop {} {\bibfield  {journal} {\bibinfo  {journal} {Sov. Phys. JETP}\
  }\textbf {\bibinfo {volume} {20}},\ \bibinfo {pages} {1018--1026} (\bibinfo
  {year} {1965})}\BibitemShut {NoStop}%
\bibitem [{\citenamefont {Kamenev}(2011)}]{kamenev2011field}%
  \BibitemOpen
  \bibfield  {author} {\bibinfo {author} {\bibfnamefont {Alex}\ \bibnamefont
  {Kamenev}},\ }\href@noop {} {\emph {\bibinfo {title} {Field theory of
  non-equilibrium systems}}}\ (\bibinfo  {publisher} {Cambridge University
  Press},\ \bibinfo {year} {2011})\BibitemShut {NoStop}%
\bibitem [{\citenamefont {Bar~Lev}\ \emph {et~al.}(2015)\citenamefont
  {Bar~Lev}, \citenamefont {Cohen},\ and\ \citenamefont
  {Reichman}}]{barLev2015}%
  \BibitemOpen
  \bibfield  {author} {\bibinfo {author} {\bibfnamefont {Yevgeny}\ \bibnamefont
  {Bar~Lev}}, \bibinfo {author} {\bibfnamefont {Guy}\ \bibnamefont {Cohen}}, \
  and\ \bibinfo {author} {\bibfnamefont {David~R.}\ \bibnamefont {Reichman}},\
  }\bibfield  {title} {\enquote {\bibinfo {title} {Absence of diffusion in an
  interacting system of spinless fermions on a one-dimensional disordered
  lattice},}\ }\href@noop {} {\bibfield  {journal} {\bibinfo  {journal} {Phys.
  Rev. Lett.}\ }\textbf {\bibinfo {volume} {114}},\ \bibinfo {pages} {100601}
  (\bibinfo {year} {2015})}\BibitemShut {NoStop}%
\bibitem [{\citenamefont {Gopalakrishnan}\ \emph {et~al.}(2016)\citenamefont
  {Gopalakrishnan}, \citenamefont {Agarwal}, \citenamefont {Demler},
  \citenamefont {Huse},\ and\ \citenamefont
  {Knap}}]{gopalakrishnan2016griffiths}%
  \BibitemOpen
  \bibfield  {author} {\bibinfo {author} {\bibfnamefont {Sarang}\ \bibnamefont
  {Gopalakrishnan}}, \bibinfo {author} {\bibfnamefont {Kartiek}\ \bibnamefont
  {Agarwal}}, \bibinfo {author} {\bibfnamefont {Eugene~A}\ \bibnamefont
  {Demler}}, \bibinfo {author} {\bibfnamefont {David~A}\ \bibnamefont {Huse}},
  \ and\ \bibinfo {author} {\bibfnamefont {Michael}\ \bibnamefont {Knap}},\
  }\bibfield  {title} {\enquote {\bibinfo {title} {Griffiths effects and slow
  dynamics in nearly many-body localized systems},}\ }\href@noop {} {\bibfield
  {journal} {\bibinfo  {journal} {Physical Review B}\ }\textbf {\bibinfo
  {volume} {93}},\ \bibinfo {pages} {134206} (\bibinfo {year}
  {2016})}\BibitemShut {NoStop}%
\bibitem [{\citenamefont {Gopalakrishnan}\ \emph {et~al.}(2015)\citenamefont
  {Gopalakrishnan}, \citenamefont {M\"uller}, \citenamefont {Khemani},
  \citenamefont {Knap}, \citenamefont {Demler},\ and\ \citenamefont
  {Huse}}]{gopalakrishnanMottCond2015}%
  \BibitemOpen
  \bibfield  {author} {\bibinfo {author} {\bibfnamefont {Sarang}\ \bibnamefont
  {Gopalakrishnan}}, \bibinfo {author} {\bibfnamefont {Markus}\ \bibnamefont
  {M\"uller}}, \bibinfo {author} {\bibfnamefont {Vedika}\ \bibnamefont
  {Khemani}}, \bibinfo {author} {\bibfnamefont {Michael}\ \bibnamefont {Knap}},
  \bibinfo {author} {\bibfnamefont {Eugene}\ \bibnamefont {Demler}}, \ and\
  \bibinfo {author} {\bibfnamefont {David~A.}\ \bibnamefont {Huse}},\
  }\bibfield  {title} {\enquote {\bibinfo {title} {Low-frequency conductivity
  in many-body localized systems},}\ }\href@noop {} {\bibfield  {journal}
  {\bibinfo  {journal} {Phys. Rev. B}\ }\textbf {\bibinfo {volume} {92}},\
  \bibinfo {pages} {104202} (\bibinfo {year} {2015})}\BibitemShut {NoStop}%
\bibitem [{\citenamefont {Mierzejewski}\ \emph {et~al.}(2016)\citenamefont
  {Mierzejewski}, \citenamefont {Herbrych},\ and\ \citenamefont
  {Prelov\ifmmode~\check{s}\else \v{s}\fi{}ek}}]{PhysRevB.94.224207}%
  \BibitemOpen
  \bibfield  {author} {\bibinfo {author} {\bibfnamefont {M.}~\bibnamefont
  {Mierzejewski}}, \bibinfo {author} {\bibfnamefont {J.}~\bibnamefont
  {Herbrych}}, \ and\ \bibinfo {author} {\bibfnamefont {P.}~\bibnamefont
  {Prelov\ifmmode~\check{s}\else \v{s}\fi{}ek}},\ }\bibfield  {title} {\enquote
  {\bibinfo {title} {Universal dynamics of density correlations at the
  transition to the many-body localized state},}\ }\href@noop {} {\bibfield
  {journal} {\bibinfo  {journal} {Phys. Rev. B}\ }\textbf {\bibinfo {volume}
  {94}},\ \bibinfo {pages} {224207} (\bibinfo {year} {2016})}\BibitemShut
  {NoStop}%
\bibitem [{\citenamefont {{Prelov\ifmmode \check{s}\else \v{s}\fi{}ek, P. and
  Mierzejewski, M. and Barišić, O. and Herbrych,
  J.}}(2017)}]{PrelovsekHerbrych17}%
  \BibitemOpen
  \bibfield  {author} {\bibinfo {author} {\bibnamefont {{Prelov\ifmmode
  \check{s}\else \v{s}\fi{}ek, P. and Mierzejewski, M. and Barišić, O. and
  Herbrych, J.}}},\ }\bibfield  {title} {\enquote {\bibinfo {title} {Density
  correlations and transport in models of many-body localization},}\
  }\href@noop {} {\bibfield  {journal} {\bibinfo  {journal} {Annalen der
  Physik}\ }\textbf {\bibinfo {volume} {529}},\ \bibinfo {pages} {1600362}
  (\bibinfo {year} {2017})}\BibitemShut {NoStop}%
\bibitem [{\citenamefont {Agarwal}\ \emph {et~al.}(2017)\citenamefont
  {Agarwal}, \citenamefont {Altman}, \citenamefont {Demler}, \citenamefont
  {Gopalakrishnan}, \citenamefont {Huse},\ and\ \citenamefont
  {Knap}}]{agarwal2017rare}%
  \BibitemOpen
  \bibfield  {author} {\bibinfo {author} {\bibfnamefont {Kartiek}\ \bibnamefont
  {Agarwal}}, \bibinfo {author} {\bibfnamefont {Ehud}\ \bibnamefont {Altman}},
  \bibinfo {author} {\bibfnamefont {Eugene}\ \bibnamefont {Demler}}, \bibinfo
  {author} {\bibfnamefont {Sarang}\ \bibnamefont {Gopalakrishnan}}, \bibinfo
  {author} {\bibfnamefont {David~A}\ \bibnamefont {Huse}}, \ and\ \bibinfo
  {author} {\bibfnamefont {Michael}\ \bibnamefont {Knap}},\ }\bibfield  {title}
  {\enquote {\bibinfo {title} {Rare-region effects and dynamics near the
  many-body localization transition},}\ }\href@noop {} {\bibfield  {journal}
  {\bibinfo  {journal} {Annalen der Physik}\ } (\bibinfo {year}
  {2017})}\BibitemShut {NoStop}%
\bibitem [{\citenamefont {{\ifmmode \check{Z}\else \v{Z}\fi{}nidari\ifmmode
  \check{c}\else \v{c}\fi{}, Marko and Scardicchio, Antonello and Varma, Vipin
  Kerala}}(2016)}]{PhysRevLett.117.040601}%
  \BibitemOpen
  \bibfield  {author} {\bibinfo {author} {\bibnamefont {{\ifmmode
  \check{Z}\else \v{Z}\fi{}nidari\ifmmode \check{c}\else \v{c}\fi{}, Marko and
  Scardicchio, Antonello and Varma, Vipin Kerala}}},\ }\bibfield  {title}
  {\enquote {\bibinfo {title} {{Diffusive and Subdiffusive Spin Transport in
  the Ergodic Phase of a Many-Body Localizable System}},}\ }\href@noop {}
  {\bibfield  {journal} {\bibinfo  {journal} {Phys. Rev. Lett.}\ }\textbf
  {\bibinfo {volume} {117}},\ \bibinfo {pages} {040601} (\bibinfo {year}
  {2016})}\BibitemShut {NoStop}%
\bibitem [{\citenamefont {Lev}\ \emph {et~al.}(2017)\citenamefont {Lev},
  \citenamefont {Kennes}, \citenamefont {Klöckner}, \citenamefont {Reichman},\
  and\ \citenamefont {Karrasch}}]{Karrasch_2017}%
  \BibitemOpen
  \bibfield  {author} {\bibinfo {author} {\bibfnamefont {Yevgeny~Bar}\
  \bibnamefont {Lev}}, \bibinfo {author} {\bibfnamefont {Dante~M.}\
  \bibnamefont {Kennes}}, \bibinfo {author} {\bibfnamefont {Christian}\
  \bibnamefont {Klöckner}}, \bibinfo {author} {\bibfnamefont {David~R.}\
  \bibnamefont {Reichman}}, \ and\ \bibinfo {author} {\bibfnamefont
  {Christoph}\ \bibnamefont {Karrasch}},\ }\bibfield  {title} {\enquote
  {\bibinfo {title} {Transport in quasiperiodic interacting systems: From
  superdiffusion to subdiffusion},}\ }\href@noop {} {\bibfield  {journal}
  {\bibinfo  {journal} {EPL (Europhysics Letters)}\ }\textbf {\bibinfo {volume}
  {119}},\ \bibinfo {pages} {37003} (\bibinfo {year} {2017})}\BibitemShut
  {NoStop}%
\bibitem [{\citenamefont {{\v Z}nidari{\v c}}\ and\ \citenamefont
  {Ljubotina}(2018)}]{znidaric2018}%
  \BibitemOpen
  \bibfield  {author} {\bibinfo {author} {\bibfnamefont {Marko}\ \bibnamefont
  {{\v Z}nidari{\v c}}}\ and\ \bibinfo {author} {\bibfnamefont {Marko}\
  \bibnamefont {Ljubotina}},\ }\bibfield  {title} {\enquote {\bibinfo {title}
  {Interaction instability of localization in quasiperiodic systems},}\
  }\href@noop {} {\bibfield  {journal} {\bibinfo  {journal} {Proceedings of the
  National Academy of Sciences}\ }\textbf {\bibinfo {volume} {115}},\ \bibinfo
  {pages} {4595--4600} (\bibinfo {year} {2018})}\BibitemShut {NoStop}%
\bibitem [{\citenamefont {Edwards}\ and\ \citenamefont
  {Anderson}(1975)}]{AndersonReplica}%
  \BibitemOpen
  \bibfield  {author} {\bibinfo {author} {\bibfnamefont {S.~F.}\ \bibnamefont
  {Edwards}}\ and\ \bibinfo {author} {\bibfnamefont {P.~W.}\ \bibnamefont
  {Anderson}},\ }\bibfield  {title} {\enquote {\bibinfo {title} {Theory of spin
  glasses},}\ }\href@noop {} {\bibfield  {journal} {\bibinfo  {journal}
  {Journal of Physics F: Metal Physics}\ }\textbf {\bibinfo {volume} {5}},\
  \bibinfo {pages} {965} (\bibinfo {year} {1975})}\BibitemShut {NoStop}%
\bibitem [{\citenamefont {Bar~Lev}\ and\ \citenamefont
  {Reichman}(2014)}]{barlev1}%
  \BibitemOpen
  \bibfield  {author} {\bibinfo {author} {\bibfnamefont {Yevgeny}\ \bibnamefont
  {Bar~Lev}}\ and\ \bibinfo {author} {\bibfnamefont {David~R.}\ \bibnamefont
  {Reichman}},\ }\bibfield  {title} {\enquote {\bibinfo {title} {Dynamics of
  many-body localization},}\ }\href@noop {} {\bibfield  {journal} {\bibinfo
  {journal} {Phys. Rev. B}\ }\textbf {\bibinfo {volume} {89}},\ \bibinfo
  {pages} {220201} (\bibinfo {year} {2014})}\BibitemShut {NoStop}%
\bibitem [{\citenamefont {Lev}\ and\ \citenamefont {Reichman}(2016)}]{barlev2}%
  \BibitemOpen
  \bibfield  {author} {\bibinfo {author} {\bibfnamefont {Yevgeny~Bar}\
  \bibnamefont {Lev}}\ and\ \bibinfo {author} {\bibfnamefont {David~R.}\
  \bibnamefont {Reichman}},\ }\bibfield  {title} {\enquote {\bibinfo {title}
  {Slow dynamics in a two-dimensional anderson-hubbard model},}\ }\href@noop {}
  {\bibfield  {journal} {\bibinfo  {journal} {EPL (Europhysics Letters)}\
  }\textbf {\bibinfo {volume} {113}},\ \bibinfo {pages} {46001} (\bibinfo
  {year} {2016})}\BibitemShut {NoStop}%
\bibitem [{Note1()}]{Note1}%
  \BibitemOpen
  \bibinfo {note} {This is only strictly valid in the thermodynamic limit, for
  a finite size system $\protect \mathcal {C}(t)$ will attain a residual value
  $\sim 1/L$ for late times even in the delocalized phase.}\BibitemShut {Stop}%
\bibitem [{\citenamefont {Aubry}\ and\ \citenamefont
  {Andr{\'e}}(1980)}]{aubry1980analyticity}%
  \BibitemOpen
  \bibfield  {author} {\bibinfo {author} {\bibfnamefont {Serge}\ \bibnamefont
  {Aubry}}\ and\ \bibinfo {author} {\bibfnamefont {Gilles}\ \bibnamefont
  {Andr{\'e}}},\ }\bibfield  {title} {\enquote {\bibinfo {title} {Analyticity
  breaking and anderson localization in incommensurate lattices},}\ }\href@noop
  {} {\bibfield  {journal} {\bibinfo  {journal} {Ann. Israel Phys. Soc}\
  }\textbf {\bibinfo {volume} {3}},\ \bibinfo {pages} {18} (\bibinfo {year}
  {1980})}\BibitemShut {NoStop}%
\bibitem [{\citenamefont {Devakul}\ and\ \citenamefont
  {Singh}(2015)}]{devakul2015}%
  \BibitemOpen
  \bibfield  {author} {\bibinfo {author} {\bibfnamefont {Trithep}\ \bibnamefont
  {Devakul}}\ and\ \bibinfo {author} {\bibfnamefont {Rajiv R.~P.}\ \bibnamefont
  {Singh}},\ }\bibfield  {title} {\enquote {\bibinfo {title} {Early breakdown
  of area-law entanglement at the many-body delocalization transition},}\
  }\href@noop {} {\bibfield  {journal} {\bibinfo  {journal} {Phys. Rev. Lett.}\
  }\textbf {\bibinfo {volume} {115}},\ \bibinfo {pages} {187201} (\bibinfo
  {year} {2015})}\BibitemShut {NoStop}%
\bibitem [{\citenamefont {{Doggen}}\ \emph {et~al.}(2018)\citenamefont
  {{Doggen}}, \citenamefont {{Schindler}}, \citenamefont {{Tikhonov}},
  \citenamefont {{Mirlin}}, \citenamefont {{Neupert}}, \citenamefont
  {{Polyakov}},\ and\ \citenamefont {{Gornyi}}}]{2018arXiv180705051D}%
  \BibitemOpen
  \bibfield  {author} {\bibinfo {author} {\bibfnamefont {E.~V.~H.}\
  \bibnamefont {{Doggen}}}, \bibinfo {author} {\bibfnamefont {F.}~\bibnamefont
  {{Schindler}}}, \bibinfo {author} {\bibfnamefont {K.~S.}\ \bibnamefont
  {{Tikhonov}}}, \bibinfo {author} {\bibfnamefont {A.~D.}\ \bibnamefont
  {{Mirlin}}}, \bibinfo {author} {\bibfnamefont {T.}~\bibnamefont {{Neupert}}},
  \bibinfo {author} {\bibfnamefont {D.~G.}\ \bibnamefont {{Polyakov}}}, \ and\
  \bibinfo {author} {\bibfnamefont {I.~V.}\ \bibnamefont {{Gornyi}}},\
  }\bibfield  {title} {\enquote {\bibinfo {title} {{Many-body (de)localization
  in large quantum chains}},}\ }\href@noop {} {\bibfield  {journal} {\bibinfo
  {journal} {arXiv:1807.05051}\ } (\bibinfo {year} {2018})}\BibitemShut
  {NoStop}%
\bibitem [{\citenamefont {Gopalakrishnan}\ \emph {et~al.}(2017)\citenamefont
  {Gopalakrishnan}, \citenamefont {Islam},\ and\ \citenamefont
  {Knap}}]{gopalakrishnan2017noise}%
  \BibitemOpen
  \bibfield  {author} {\bibinfo {author} {\bibfnamefont {Sarang}\ \bibnamefont
  {Gopalakrishnan}}, \bibinfo {author} {\bibfnamefont {K.~Ranjibul}\
  \bibnamefont {Islam}}, \ and\ \bibinfo {author} {\bibfnamefont {Michael}\
  \bibnamefont {Knap}},\ }\bibfield  {title} {\enquote {\bibinfo {title}
  {Noise-induced subdiffusion in strongly localized quantum systems},}\
  }\href@noop {} {\bibfield  {journal} {\bibinfo  {journal} {Physical Review
  Letters}\ }\textbf {\bibinfo {volume} {119}},\ \bibinfo {pages} {046601}
  (\bibinfo {year} {2017})}\BibitemShut {NoStop}%
\bibitem [{\citenamefont {Nandkishore}\ \emph {et~al.}(2014)\citenamefont
  {Nandkishore}, \citenamefont {Gopalakrishnan},\ and\ \citenamefont
  {Huse}}]{nandkishore2014spectral}%
  \BibitemOpen
  \bibfield  {author} {\bibinfo {author} {\bibfnamefont {Rahul}\ \bibnamefont
  {Nandkishore}}, \bibinfo {author} {\bibfnamefont {Sarang}\ \bibnamefont
  {Gopalakrishnan}}, \ and\ \bibinfo {author} {\bibfnamefont {David~A.}\
  \bibnamefont {Huse}},\ }\bibfield  {title} {\enquote {\bibinfo {title}
  {Spectral features of a many-body-localized system weakly coupled to a
  bath},}\ }\href@noop {} {\bibfield  {journal} {\bibinfo  {journal} {Physical
  Review B}\ }\textbf {\bibinfo {volume} {90}},\ \bibinfo {pages} {064203}
  (\bibinfo {year} {2014})}\BibitemShut {NoStop}%
\bibitem [{\citenamefont {{Jonathan Wurtz and Anatoli Polkovnikov and Dries
  Sels}}(2018)}]{WURTZ2018341}%
  \BibitemOpen
  \bibfield  {author} {\bibinfo {author} {\bibnamefont {{Jonathan Wurtz and
  Anatoli Polkovnikov and Dries Sels}}},\ }\bibfield  {title} {\enquote
  {\bibinfo {title} {{Cluster truncated Wigner approximation in strongly
  interacting systems}},}\ }\href@noop {} {\bibfield  {journal} {\bibinfo
  {journal} {Annals of Physics}\ }\textbf {\bibinfo {volume} {395}},\ \bibinfo
  {pages} {341 -- 365} (\bibinfo {year} {2018})}\BibitemShut {NoStop}%
\bibitem [{\citenamefont {Moses}\ \emph {et~al.}(2017)\citenamefont {Moses},
  \citenamefont {Covey}, \citenamefont {Miecnikowski}, \citenamefont {Jin},\
  and\ \citenamefont {Ye}}]{moses_new_2017}%
  \BibitemOpen
  \bibfield  {author} {\bibinfo {author} {\bibfnamefont {Steven~A.}\
  \bibnamefont {Moses}}, \bibinfo {author} {\bibfnamefont {Jacob~P.}\
  \bibnamefont {Covey}}, \bibinfo {author} {\bibfnamefont {Matthew~T.}\
  \bibnamefont {Miecnikowski}}, \bibinfo {author} {\bibfnamefont {Deborah~S.}\
  \bibnamefont {Jin}}, \ and\ \bibinfo {author} {\bibfnamefont {Jun}\
  \bibnamefont {Ye}},\ }\bibfield  {title} {\enquote {\bibinfo {title} {New
  frontiers for quantum gases of polar molecules},}\ }\href@noop {} {\bibfield
  {journal} {\bibinfo  {journal} {Nature Physics}\ }\textbf {\bibinfo {volume}
  {13}},\ \bibinfo {pages} {13--20} (\bibinfo {year} {2017})}\BibitemShut
  {NoStop}%
\bibitem [{\citenamefont {Bohrdt}\ \emph {et~al.}(2018)\citenamefont {Bohrdt},
  \citenamefont {Greif}, \citenamefont {Demler}, \citenamefont {Knap},\ and\
  \citenamefont {Grusdt}}]{bohrdt_angle-resolved_2018}%
  \BibitemOpen
  \bibfield  {author} {\bibinfo {author} {\bibfnamefont {A.}~\bibnamefont
  {Bohrdt}}, \bibinfo {author} {\bibfnamefont {D.}~\bibnamefont {Greif}},
  \bibinfo {author} {\bibfnamefont {E.}~\bibnamefont {Demler}}, \bibinfo
  {author} {\bibfnamefont {M.}~\bibnamefont {Knap}}, \ and\ \bibinfo {author}
  {\bibfnamefont {F.}~\bibnamefont {Grusdt}},\ }\bibfield  {title} {\enquote
  {\bibinfo {title} {Angle-resolved photoemission spectroscopy with quantum gas
  microscopes},}\ }\href@noop {} {\bibfield  {journal} {\bibinfo  {journal}
  {Phys. Rev. B}\ }\textbf {\bibinfo {volume} {97}},\ \bibinfo {pages} {125117}
  (\bibinfo {year} {2018})}\BibitemShut {NoStop}%
\bibitem [{\citenamefont {{Aarts, Gert and Berges,
  J\"urgen}}(2001)}]{AartsBerges01}%
  \BibitemOpen
  \bibfield  {author} {\bibinfo {author} {\bibnamefont {{Aarts, Gert and
  Berges, J\"urgen}}},\ }\bibfield  {title} {\enquote {\bibinfo {title}
  {Nonequilibrium time evolution of the spectral function in quantum field
  theory},}\ }\href@noop {} {\bibfield  {journal} {\bibinfo  {journal} {Phys.
  Rev. D}\ }\textbf {\bibinfo {volume} {64}},\ \bibinfo {pages} {105010}
  (\bibinfo {year} {2001})}\BibitemShut {NoStop}%
\bibitem [{\citenamefont {{J\"urgen Berges and J\"urgen
  Cox}}(2001)}]{BERGES2001369}%
  \BibitemOpen
  \bibfield  {author} {\bibinfo {author} {\bibnamefont {{J\"urgen Berges and
  J\"urgen Cox}}},\ }\bibfield  {title} {\enquote {\bibinfo {title}
  {Thermalization of quantum fields from time-reversal invariant evolution
  equations},}\ }\href@noop {} {\bibfield  {journal} {\bibinfo  {journal}
  {Physics Letters B}\ }\textbf {\bibinfo {volume} {517}},\ \bibinfo {pages}
  {369 -- 374} (\bibinfo {year} {2001})}\BibitemShut {NoStop}%
\bibitem [{\citenamefont {{J\"urgen Berges}}(2002)}]{BERGES2002847}%
  \BibitemOpen
  \bibfield  {author} {\bibinfo {author} {\bibnamefont {{J\"urgen Berges}}},\
  }\bibfield  {title} {\enquote {\bibinfo {title} {Controlled nonperturbative
  dynamics of quantum fields out of equilibrium},}\ }\href@noop {} {\bibfield
  {journal} {\bibinfo  {journal} {Nuclear Physics A}\ }\textbf {\bibinfo
  {volume} {699}},\ \bibinfo {pages} {847 -- 886} (\bibinfo {year}
  {2002})}\BibitemShut {NoStop}%
\bibitem [{\citenamefont {Weidinger}\ and\ \citenamefont
  {Knap}(2017)}]{Weidinger17}%
  \BibitemOpen
  \bibfield  {author} {\bibinfo {author} {\bibfnamefont {Simon~A.}\
  \bibnamefont {Weidinger}}\ and\ \bibinfo {author} {\bibfnamefont {Michael}\
  \bibnamefont {Knap}},\ }\bibfield  {title} {\enquote {\bibinfo {title}
  {{Floquet prethermalization and regimes of heating in a periodically driven,
  interacting quantum system}},}\ }\href@noop {} {\bibfield  {journal}
  {\bibinfo  {journal} {Sci. Rep.}\ }\textbf {\bibinfo {volume} {7}},\ \bibinfo
  {pages} {45382} (\bibinfo {year} {2017})}\BibitemShut {NoStop}%
\end{thebibliography}%
\end{document}